\documentclass[reprint, amsmath, amssymb, aps, pra]{revtex4-2}
\usepackage{graphicx}
\usepackage{dcolumn}
\usepackage{bm}
\usepackage{braket}
\usepackage{siunitx}
\usepackage{physics}
\usepackage[colorlinks]{hyperref}

\hypersetup{allcolors=blue}
\usepackage{changes}
\usepackage[ruled,vlined,linesnumbered]{algorithm2e}
\usepackage{algorithmic}

\usepackage{amsthm}
\newtheorem{theorem}{Theorem}

\newtheorem{problem}{Problem}

\usepackage{cleveref}
\crefname{problem}{Problem}{Problems}

\newcommand{\thm}[1]{\hyperref[thm:#1]{Theorem~\ref*{thm:#1}}}
\newcommand{\cor}[1]{\hyperref[cor:#1]{Corollary~\ref*{cor:#1}}}
\newcommand{\defn}[1]{\hyperref[defn:#1]{Definition~\ref*{defn:#1}}}
\newcommand{\lem}[1]{\hyperref[lem:#1]{Lemma~\ref*{lem:#1}}}
\newcommand{\prop}[1]{\hyperref[prop:#1]{Proposition~\ref*{prop:#1}}}
\newcommand{\assum}[1]{\hyperref[assum:#1]{Assumption~\ref*{assum:#1}}}
\newcommand{\fig}[2][]{\hyperref[fig:#2]{Figure~\ref*{fig:#2}#1}}
\newcommand{\tab}[1]{\hyperref[tab:#1]{Table~\ref*{tab:#1}}}
\newcommand{\algo}[1]{\hyperref[algo:#1]{Algorithm~\ref*{algo:#1}}}
\newcommand{\app}[1]{\hyperref[app:#1]{Supplementary Note~\ref*{app:#1}}}
\newcommand{\prob}[1]{\hyperref[Problem:#1]{Problem~\ref*{Problem:#1}}}

\DeclareRobustCommand{\red}[1]{{\color{red}\hypersetup{allcolors=red}#1}}
\renewcommand{\red}[1]{#1}

\begin{document}
\preprint{APS/123-QED}
\title{Non-Convex Optimization via Quantum Tunneling Search on a Photonic Chip}
\author{Xi-Jun Yuan$^{1,2}$}
\thanks{Equal Contribution.}

\author{Shuo Zhou$^{3}$}
\thanks{Equal Contribution.}

\author{Yizhou Liu$^{4}$}

\author{Xuan Chen$^{1,2}$}
\author{Hong-Jing Du$^{1,2}$}
\author{Xiao-Wen Shang$^{1,2}$}
\author{Wen-Hao Zhou$^{1,2}$}
\author{Yu-Xuan Fu$^{1,2}$}
\author{Bao-Jing Liu$^{1,2}$}
\author{Ze-Feng Lan$^{1,2}$}
\author{Hang Song$^{1,2}$}
\author{Fang-Song Nie$^{5}$}
\author{Yu-Lin Yu$^{5}$}
\author{Xiao-Yun Xu$^{1,2,6}$}
\author{Jian-Peng Dou$^{1,2}$}

\author{Tongyang Li$^{3}$}
\thanks{Corresponding author. Email: {\hypersetup{urlcolor=magenta}\href{mailto:tongyangli@pku.edu.cn}{tongyangli@pku.edu.cn}}}

\author{Hao Tang$^{1,2}$}
\thanks{Corresponding author. Email: {\hypersetup{urlcolor=magenta}\href{mailto:htang2015@sjtu.edu.cn}{htang2015@sjtu.edu.cn}}}

\author{Xian-Min Jin$^{1,2,5,6}$}
\thanks{Corresponding author. Email: {\hypersetup{urlcolor=magenta}\href{mailto:xianmin.jin@sjtu.edu.cn}{xianmin.jin@sjtu.edu.cn}}}

\affiliation{\vspace{1em}\textsuperscript{1}Center for Integrated Quantum Information Technologies (IQIT), School of Physics and Astronomy and State Key Laboratory of Advanced Optical Communication Systems and Networks, Shanghai Jiao Tong University, Shanghai 200240, China}

\affiliation{\textsuperscript{2}Hefei National Laboratory, Hefei 230088, China}

\affiliation{\textsuperscript{3}Center on Frontiers of Computing Studies, School of Computer Science, Peking University, Beijing 100871, China
}
\affiliation{\textsuperscript{4}Department of Engineering Mechanics, Tsinghua University, Beijing 100084, China
}

\affiliation{\textsuperscript{5}TuringQ Company, Ltd., Shanghai 200240, China}

\affiliation{\textsuperscript{6}Chip Hub for Integrated Photonics Xplore (CHIPX), Shanghai Jiao Tong University, Wuxi 214000, China}



\begin{abstract}
Non-convex optimization plays an important role in modern machine learning since plenty of real-world problems have objective functions or constraints with non-convex properties. The non-convex nature of landscapes gives rise to numerous local minima and saddle points, dramatically hindering the ability of gradient descent-based methods to explore the landscape and obtain global optima. Here we present quantum tunneling search (QTS), an algorithm based on quantum tunneling walk and local annealing, which effectively explores the non-convex landscape by leveraging the quantum tunneling effect. In experiments, we map an objective function to the potential term of the waveguide Hamiltonian by controlling the refractive index with the femtosecond direct laser writing technique. With a proper input state, the average probability distributions on $\SI{10}{\micro\meter}$-spaced 29$\times$1 and 21$\times$9 waveguide arrays highlight the local minima of various 1D and 2D non-convex landscapes, respectively. By further introducing the varying Hamiltonian approach, more efficient search and even the identification of the global minimum are achieved. Our work experimentally verifies that QTS is a successful and efficient non-convex optimization solver, demonstrating quantum speedups over classical stochastic gradient descent (SGD), which serves as a practical application of near-term analog quantum simulators in the field of machine learning.
\end{abstract}

\maketitle

\section{\label{sec:level1} INTRODUCTION}

Recent advances in machine learning, especially the applications of neural networks, solicit the study of non-convex optimization. This is because the loss function of a deep neural network with a huge number of parameters is intrinsically high-dimensional, which often exhibits complex landscapes characterized by many local minima and intricate saddle points, in stark contrast to the simpler convex landscapes. Those non-convex optimization problems are notorious for being NP-hard~\cite{jain2017non, Goodfellow-et-al-2016}, and there are even instances where finding approximation solutions is also NP-hard~\cite{meka2008rank}.

The gradient descent method is ubiquitous for solving optimization problems, but it suffers from gradient vanishing in non-convex landscapes, making it prone to getting trapped in local minima~\cite{glorot2010understanding}. On the other hand, as quantum evolution involves wave properties, particles can probabilistically tunnel through energy barriers and barren regions even when their energy is lower than the height of the barrier, which is known as the \emph{quantum tunneling} effect. This property illuminates the idea of employing quantum evolution for non-convex optimization. A former study~\cite{Liu_2023} proposed quantum tunneling walk (QTW), which exhibits faster search speeds compared to stochastic gradient descent (SGD). In addition, there are research endeavors exploring the utilization of quantum evolution as a replacement for gradient descent~\cite{Rebentrost_2019,leng2023quantumhamiltoniandescent,Chen_2025}. Beyond classical objective functions, Ref.~\cite{chen_local_2025} demonstrated quantum speedups in finding local minima of quantum systems. However, few of them have been experimentally verified on actual quantum devices.

\begin{figure*}[!t]
  \centering
  \includegraphics[width=\linewidth]{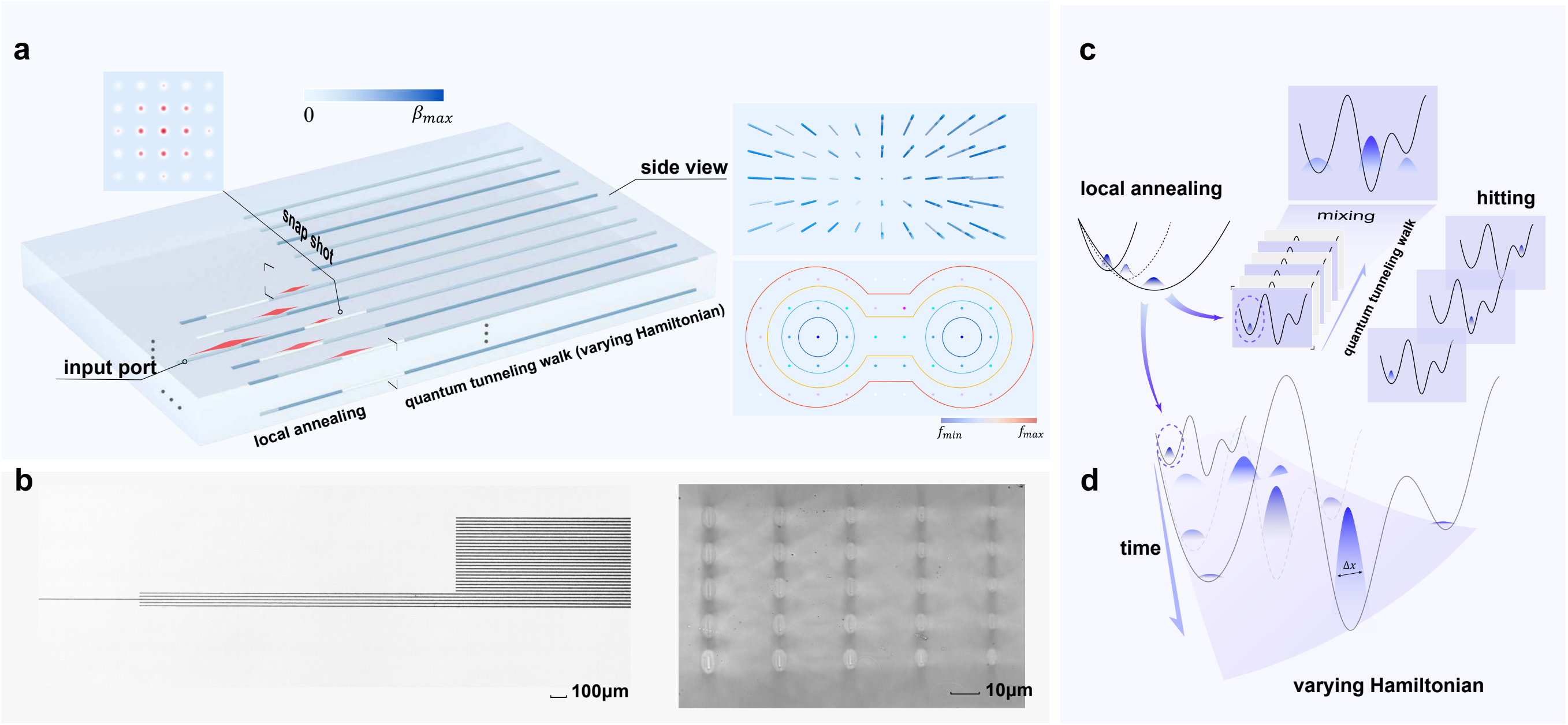}
  \caption{\textbf{Diagram of our photonic chip and quantum tunneling search.} \textbf{a} Schematic diagram of a three-dimensional photonic chip prepared by laser direct writing, with different colors representing the degree of change in the refractive index of the waveguide, as measured by $\beta$. The red line shows the propagation of photons among the waveguides, and the snapshot gives the probability distribution of the input state after local annealing. The side view gives the variation of the refractive index in different waveguides and different positions at the same waveguide, and the picture below represents a contour plot of the potential energy function mapped to the refractive index, where the colormap corresponds to the magnitude of potential energy. {\textbf{b}} A top view and a cross-section view of the photonic chip. {\textbf{c}} Illustration of local annealing together with the quantum tunneling walk process, where the annealing occurs in part of the landscape. Here, the solid line represents the non-convex landscape mapped onto the potential energy, and the wave packet denotes the probability distribution of the quantum state. The state after annealing serves as the input state for the whole picture, which is marked out with a dotted circle. The input state completes the search process by QTW and averaging the probability over evolution. Three critical moments signifying the input state hits the local minima on the right. {\textbf{d}} Illustration of the varying Hamiltonian approach, a method we use to strengthen the quantum tunneling walk process. By modifying the ratio of the potential term, which is represented as an increase in amplitude, the quantum state will be further localized corresponding to the lowest energy, accompanied by an increase in $\Delta x$ accuracy, enabling the search for the global minimum.}
  \label{fig:1}
\end{figure*}

Here, we present quantum tunneling search (QTS), an algorithm that works by performing quantum tunneling walk~\cite{Liu_2023} with a proper input state and then averaging the probability distribution over evolution. First, we map a non-convex function to the potential term of the waveguide Hamiltonian with the femtosecond direct laser writing technique. Next, by performing an incomplete local annealing process, we prepare a proper input state for the subsequent search process. In addition, we introduce the varying Hamiltonian approach to further accelerate the search efficiency and identify the global minimum. As a result, we experimentally verify that QTS succeeds in finding local minima and the global minimum on various 1D and 2D non-convex landscapes, demonstrating quantum speedups over classical stochastic gradient descent.

\section{Results}
\textbf{Hamiltonian mapping.} In waveguide arrays, mutually coupled waveguides can be treated as a quantum system. Throughout this work, we focus on nearest-neighbor coupling. Photons propagating through the coupled waveguide arrays can be described by the Hamiltonian
\begin{equation}
  H=\sum_i^N\beta_i a_i^\dagger a_i+\sum_{i\neq j}^NC_{i,j}a_i^\dagger a_j,
  \label{eq:Hamiltonian}
\end{equation}
where $\beta_i$ depicts the propagation constant in waveguide $i$. Here we embed the objective function into this photonic chip by setting $\beta_i$ proportional to the corresponding grid point value $f(x_i)$, as shown in \fig[a]{1}. $C_{i, j}$ is the coupling strength between the waveguides $i$ and $j$, corresponding to the kinetic energy to hop among different waveguides (see \app{details} and Ref.~\cite{wang2022topologically} for more discussion). $a_i^\dagger$, $a_i$ and $a_i^\dagger a_i$ are respectively the creation, annihilation and particle number operators of photons.

Throughout this work, we alternately adopt the propagation length $z$ to characterize quantum evolution, which can be easily controlled by setting up waveguides of certain lengths, as shown in \fig[b]{1}. In this unit system, the reduced Planck constant $\hbar = 1$, while the mapped Hamiltonian has the same dimension \SI{}{mm^{-1}} as $\beta$. Accordingly, the time evolution should be modified as
\begin{equation}
|\Psi(z)\rangle=e^{-iHz}|\Psi(0)\rangle, 
\end{equation}
where $|\Psi(0)\rangle=\Psi_{\text{in}}$ \red{and $|\Psi(z)\rangle=\sum_j c_j(z)|j\rangle$. The probability of finding the walker in waveguide $j$ after propagation length $z$ is}
\red{
\begin{equation}
    P_j(z)=|c_j(z)|^2=|\langle j|\Psi(z)\rangle|^2 .
\end{equation}
}
\red{Each waveguide represents a grid point $x_j$, with lower objective values $f(x_j)$ encoded as lower on-site potentials $\beta_j$. Peaks in the measured position probability distribution thus identify candidate minima on the grid.}

\begin{figure}[!t]
   \includegraphics[width=\linewidth]{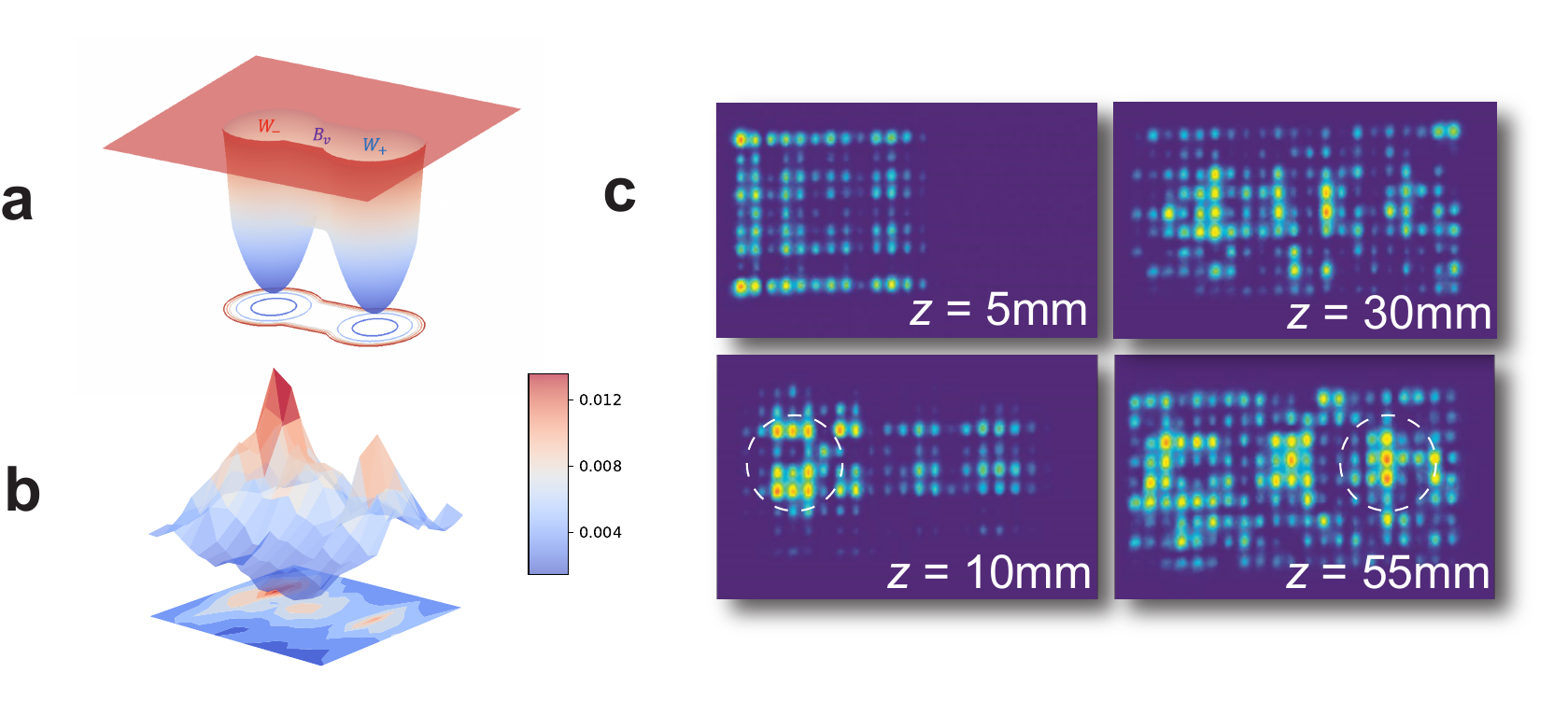}
\caption{\textbf{Double-well landscape quantum tunneling search.} \textbf{a} Illustration of the 2D landscape modified from fig.~3 in~\cite{Liu_2023}.
\textbf{b} Average probability distribution obtained experimentally from \SI{5}{mm} to \SI{90}{mm} with sampling interval of \SI{5}{mm}. \textbf{c} Snapshots of the evolutionary results obtained experimentally at $z= \SI{5}{mm},\SI{10}{mm},\SI{30}{mm}$ and \SI{55}{mm}, including the state spreading across the left well and the state hitting the right potential well, both marked out with a white circle.}
\label{fig:2}
\end{figure}

\vspace{.5em}
\textbf{Local annealing.} Quantum annealing (QA) utilizes quantum fluctuations to gradually drive the system from an arbitrary superposition to the ground state \cite{kadowaki1998quantum}, that is, the global optimal solution~\cite{jiang2018quantum,venturelli2019reverse,yuan2023quantum,mott2017solving}. However, due to noise in real quantum devices and the hardness of realizing the adiabatic condition, it is usually impractical to use QA to obtain an exact ground state. Here, we introduce an incomplete local annealing process~\cite{Grabarits_2025}.

The system evolves according to the time-dependent Schr\"{o}dinger equation
$i\hbar \frac{\partial \psi}{\partial t}=H(t)\psi$, where
\begin{equation}\label{eq:annealing}
    H(t)=\left(1-\frac{t}{T}\right)H_{\text{initial}}+\frac{t}{T}H_{\text{final}}.
\end{equation}

$H_{\text{initial}}$ is set to only part of the system at a rough local minimum, and thus we call it local annealing. As shown in \fig[c]{1}, by gradually decreasing $\beta$ in Eq.~\eqref{eq:Hamiltonian}, the potential well turns shallower, and the state spreads out. The state after local annealing will be injected into the entire non-convex landscape for the subsequent QTW to acquire the remaining local minima.

In practice, we alternatively adopt a staged incomplete local annealing approach that performs only a few steps. This yields a low-energy state whose energy is slightly higher than the ground energy of both the injection potential well and the entire landscape, which not only manages to cut down the whole infinite process~\cite{isakov2016understanding} but also produces a state with faster search speed. Similar ideas can also be found in other ground state preparation methods such as dissipative dynamics~\cite{zhan2025rapidquantumgroundstate}.

\vspace{.5em}
\textbf{Varying Hamiltonian.} In the following, we introduce the varying Hamiltonian approach to further accelerate search efficiency and identify the global minimum. During QTW, we expect a gradual increase in the depth of the potential to locate the target position with more accuracy, as shown in \fig[d]{1}. Similar ideas can also be found in other methods, such as quantum Hamiltonian descent~\cite{leng2023quantumhamiltoniandescent}, experimental Haar random \cite{Tang_2022}, reverse annealing~\cite{ohkuwa2018reverse}, and simulating photosynthetic energy transport~\cite{tang2024simulating}. \red{More broadly, engineered gauge fields provide another route for sculpting photonic modes and effective Hamiltonians on a chip~\cite{Lin2026ArtificialGaugeFields}.}
The system undergoes a process according to
\begin{equation}
    H(z) =  {C(z)}\left(-\frac{1}{2} \Delta\right) + {\beta(z)} f(x),
\end{equation}
where $\Delta$ is the Laplacian operator of the kinetic term. Experimentally, we gradually increase the value of $\beta_i$ while keeping the value of $C_{ij}$ constant.

\begin{figure}[!t]
  \includegraphics[width=\linewidth]{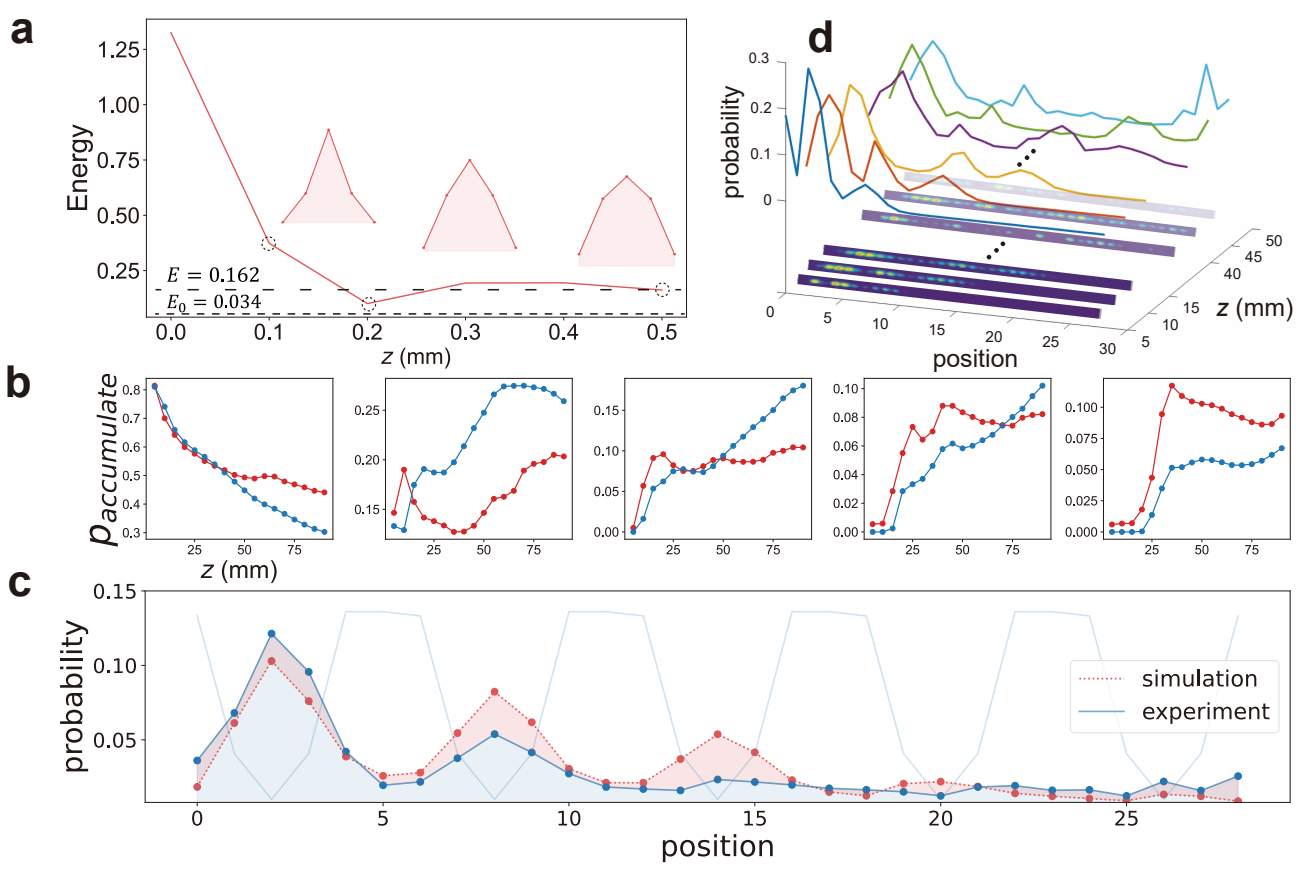}
    \caption{\textbf{Multi-well landscape quantum tunneling search.} Here we use red for simulated results and blue for experimental results.
    \textbf{a} Local annealing in the leftmost well, where $\beta_i$ varies according to $2^{6-10*z}$ scaling during each $\SI{0.1}{mm}$ stage. The energy of the input state $E=0.162$ is slightly higher than that of the ground state $E_0=0.034$ and below that of the first excited state. \textbf{b} Accumulated probability in each well and the corresponding simulated result, which is calculated by $p_{accumulate}(n\Delta z)=\sum_{i\leq n}p(i\Delta z)/n$ with sampling interval $\Delta z=\SI{5}{mm}$. \textbf{c} Average probability distribution of simulation and experiment. \textbf{d} Snapshots of the evolutionary results obtained experimentally at $z=\SI{5}{mm}, \SI{10}{mm}, \SI{15}{mm}, \SI{40}{mm}, \SI{45}{mm}$ and $\SI{50}{mm}$.}
    \label{fig:3}
\end{figure}

In the quantum Hamiltonian descent model of Leng et al.~\cite{leng2023quantumhamiltoniandescent}, the expected objective value approaches the global minimum under the assumptions of their analysis. Motivated by this result, we increase the potential strength relative to the coupling along the waveguide array and examine how the output intensity is distributed among the wells.

\vspace{.5em}
\textbf{Double-well landscape.} In experiments, we begin with the 2D landscape consisting of two potential wells and a barren region in the middle from Ref.~\cite{Liu_2023}, as shown in \fig[a]{2}. The constructed function $f$ is given by 
\begin{equation}
\left.f(\mathbf{x})=\left\{\begin{array}{l}\frac{1}{2}\omega^2\|\mathbf{x}\|^2,\:\mathbf{x}\in W_-,\\\frac{1}{2}\omega^2\|\mathbf{x}-2b\mathbf{v}\|^2,\:\mathbf{x}\in W_+,\\M_1,\:\mathbf{x}\in B_\mathbf{v},\\M_2,\:\mathrm{otherwise}.\end{array}\right.\right.
\end{equation}

We present the evolutionary results obtained experimentally on $21\times 9$ waveguide arrays in \fig[c]{2}. Here $z=\SI{55}{mm}$ denotes the hitting ``time'', that is, the minimum propagation length needed for hitting the region of interest once. Even with a non-ideal input state (not the exact ground state of the left potential well), after mixing ``time'' $z=\SI{90}{mm}$, the average probability distribution of this low-energy state still highlights both local minima, as shown in \fig[b]{2}.

\begin{figure}[!t]
  \includegraphics[width=\linewidth]{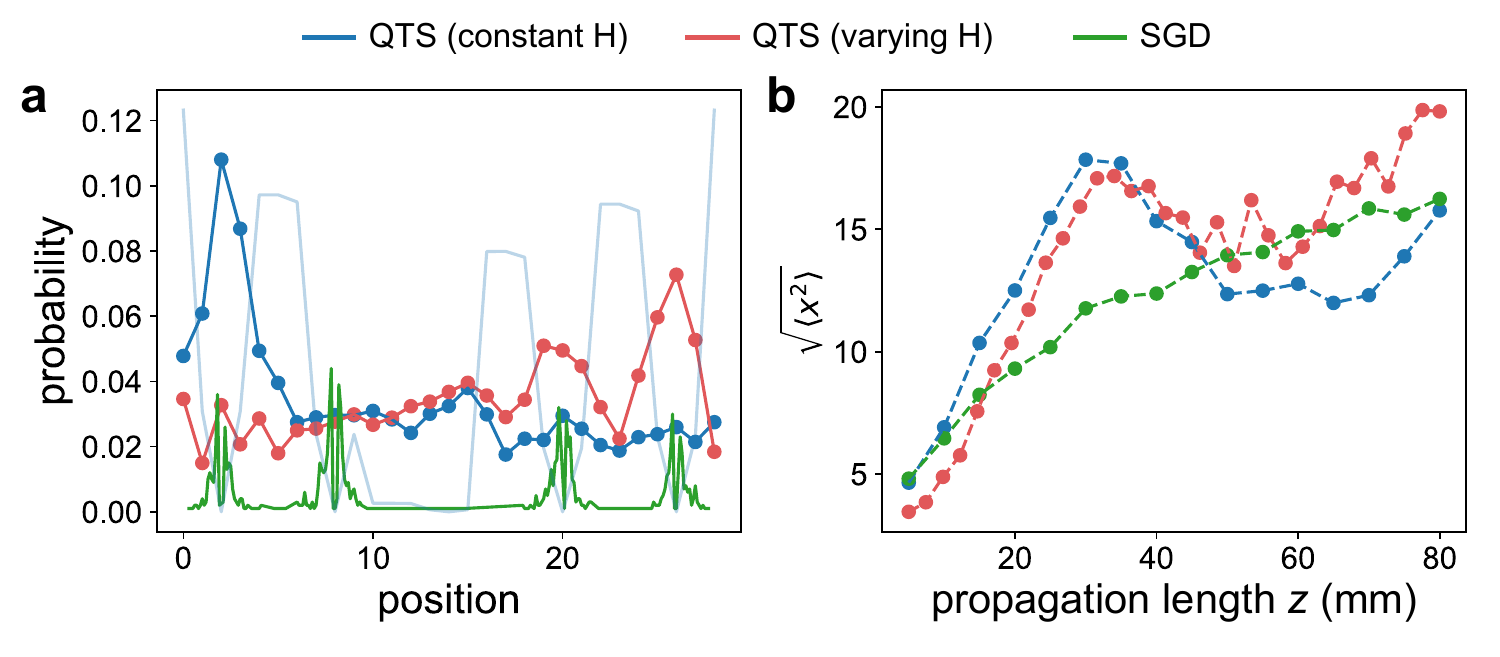}
\caption{\textbf{Multi-well random landscape quantum tunneling search.}
\textbf{a} \red{Normalized output intensities give the position distributions for QTS with a constant or varying Hamiltonian on the same five-well landscape. The constant-$H$ curve averages the sampled distributions through \SI{80}{mm}. The varying-$H$ curve shows the arithmetic mean of the last eight measured distributions. The stochastic gradient descent (SGD) curve shows the empirical distribution of positions after 1,600 iterations.}
\textbf{b} \red{Root-mean-square position $\sqrt{\langle x^2\rangle}$ versus propagation length, with SGD iterations rescaled as described in the text.}}
\label{fig:4}
\end{figure}

\vspace{.5em}
\textbf{Multi-well landscapes.} For fine-grained analysis, we conduct experiments of 1D {even} three and five-well landscapes on $29\times 1$  waveguide arrays. In the experiment, the input state is prepared by five stages of local annealing as shown in \fig[a]{3}. In simulation, we set the waveguide coupling coefficient $C = 0.47$ and apply simulated perturbations to $\beta$. Please refer to the \app{details} for error analysis. We also present the case of a three-well landscape there.

In the case of a five-well landscape, we present the evolutionary results obtained experimentally in \fig[c]{3}. As shown in \fig[c]{3}, the average probability distribution aligns well with the simulation.
In \fig[b]{3}, we observe the accumulated probability in each well, which is closely related to the mixing time. We characterize the convergence behavior of the probability distribution, and find that the maximum available waveguide length is insufficient for convergence. In the following, we alternatively employ a varying Hamiltonian to address this limitation.

\vspace{.5em}
\textbf{Multi-well random landscape.} 
To further explore the search ability, we apply the varying Hamiltonian approach to a more complex five-well landscape whose depth and width are randomly processed. For quantum tunneling search with constant Hamiltonian (QTS), we average over $z=\SI{80}{mm}$ with a sampling interval of $\SI{5}{mm}$. For quantum tunneling search with varying Hamiltonian (QTS$_t$), we take the form $H(z)=C(-\frac{1}{2} \Delta) + (z+z^3)f(x)$. 

It can be seen from \fig[a]{4} that, for a limited propagation length, both experimental approaches highlight all the local minima of the random landscape. Compared to the constant Hamiltonian, the varying Hamiltonian approach enhances the exploration of distant regions and yields a slightly more uniform probability distribution. We also implement stochastic gradient descent (SGD) \red{as a classical baseline. We sample the initial position of each SGD trajectory from a triangular distribution in the injection well and use the resulting ensemble of trajectories to estimate the position distribution. To compare the subsequent exploration behavior with propagation dynamics, the number of SGD steps is rescaled according to the propagation length divided by the learning rate, with an additional factor of 60 used only to bring its spatial spread close to that of the photonic search for an illustrative comparison. We therefore use SGD here as a baseline for comparing landscape-exploration behavior. The theoretical comparison in terms of query complexity and mixing time is discussed in \app{theory}.}

\vspace{.5em}
\textbf{Global minimum search.} Finally, we demonstrate that our experimental approach has the capability to distinguish the global optima. Here we perform quantum tunneling search with a varying Hamiltonian on a landscape with several uneven local minima beyond~\cite{Liu_2023}. As shown in \fig[a]{5}, with $H(z)=C(-\frac{1}{2} \Delta) + (0.5z+0.05z^2+0.2z^3)f(x)$, the probability near the global minimum significantly outperforms those of other local minima at the end stage $z=\SI{90}{mm}$. At each sampled propagation length $z$, \fig[b]{5} shows the fraction of output intensity in the waveguide at the global minimum ($x=14$). We also implement stochastic gradient descent as a classical baseline to demonstrate the quantum speedups in global optimal search.

\begin{figure}[!t]
  \includegraphics[width=\linewidth]{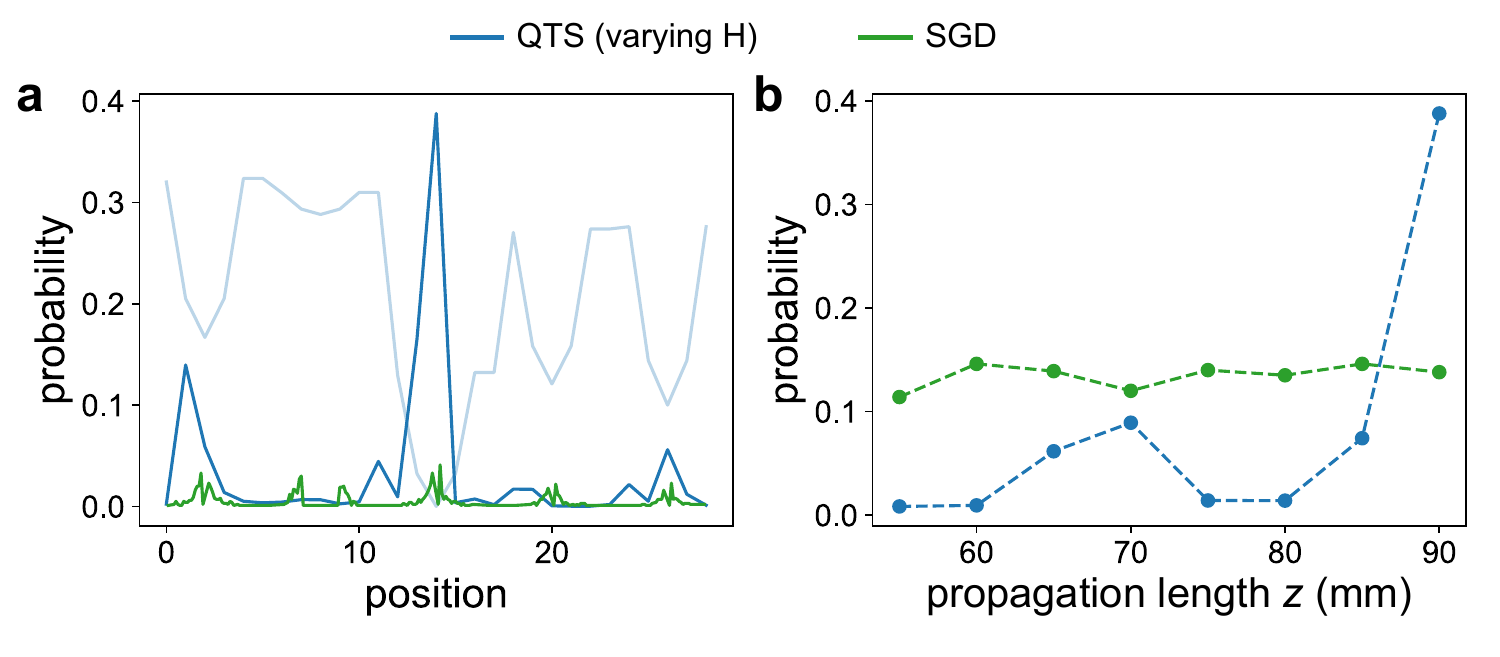}
\caption{\textbf{Global minimum search in random landscape with uneven local minima.}
\textbf{a} \red{At $z=\SI{90}{mm}$, the measured position probability for QTS with a varying Hamiltonian peaks at the global minimum ($x=14$). The SGD curve shows the endpoint distribution after 1,800 iterations on the same landscape.}
\textbf{b} \red{Probability at the global minimum versus propagation length $z$. Each QTS point gives the fraction of output intensity in the waveguide at $x=14$. Each SGD point gives the fraction of trajectories with $|x-14|<0.5$ at the corresponding rescaled iteration count.}}
\label{fig:5}
\end{figure}

\vspace{.5em}
\textbf{Future work.}
\red{Several important directions remain for future investigation. First, extending quantum tunneling search to higher-dimensional landscapes will require more compact ways to encode the objective function. In our waveguide array, each grid point occupies one waveguide, so adding a coordinate multiplies the number of waveguides by the number of grid points along that coordinate. Objectives with additive structure~\cite{kandasamy2015high} or low effective dimension~\cite{wang2013bayesian} offer natural test cases for reduced encodings. Hybrid optical and digital architectures~\cite{mcmahon2016fully} also merit study when the full grid is too large to realize on a chip.}

\red{Second, a reconfigurable chip could encode a family of objective functions on the same device~\cite{wang2025reconfigurable}. In the present chip, fabrication fixes both the couplings between waveguides and their refractive-index profiles along the propagation direction. Candidate platforms for reconfigurable encoding include circuits with thermo-optic~\cite{cao2026programmable} or electro-optic tuning~\cite{zheng2023electro}, phase-change devices~\cite{wang2016optically}, and programmable photonic meshes~\cite{perezlopez2020multipurpose}.}

\red{Third, a two-photon experiment could extend the present coherent-light study of single-particle tunneling dynamics by testing how quantum interference changes the probability of detecting both photons near the same minimum. Measuring output coincidences for distinguishable and indistinguishable photon pairs would directly test the bunching predicted in \app{open} (\fig{A10}).}

\section{Conclusion}
In this work, we introduce a novel approach for solving non-convex optimization problems by leveraging the quantum tunneling effect. We develop an efficient, scalable, and programmable non-convex optimization solver implemented on a photonic chip. In experiments, we verify that QTS succeeds in finding the local minima and the global minimum on various 1D and 2D non-convex landscapes. Given the NP-hard nature of these problems, new computational paradigms are crucial to tackling them. 


\red{As non-convex optimization remains a fundamental challenge in machine learning and other data-driven fields, photonic analog computing offers a complementary route to conventional digital and gradient-based methods. More broadly, our results suggest that photonic Hamiltonian engineering can serve as a versatile physical platform for waveguide-based optimization, opening new opportunities for optical computing beyond conventional simulation-oriented applications~\cite{wu2025monolithically,fu2026passive,yang2024programmable}.}

\begin{acknowledgments}
This research was supported by the National Key R$\&$D Program of China (Grant Numbers 2019YFA0308700, \red{2024YFA1409300}); National Natural Science Foundation of China (NSFC) (Grant Numbers 92365117, 62372006, 62235012, 12304342, \red{12574549, 12574542, 125B1033}); \red{Innovation Program for Quantum Science and Technology (Grant Numbers 2021ZD0301500, 2021ZD0300700)}; Science and Technology Commission of Shanghai Municipality (STCSM) (Grant Numbers 2019SHZDZX01, \red{24ZR1438700, 24ZR1430700, 24LZ1401500}); Startup Fund for Young Faculty at SJTU (SFYF at SJTU) (Grant Numbers \red{24X010502876}, 24X010500170); The Fundamental Research Funds for the Central Universities, Peking University. \red{Frontier Technologies R$\&$D Program of Jiangsu (Grant No. SBF20250000094).} X.-M. J. acknowledges additional support from a Shanghai talent program and support from Zhiyuan Innovative Research Center of Shanghai Jiao Tong University. \red{H. T. acknowledges additional support from Yangyang Development Fund.}
\end{acknowledgments}

\bibliography{references}

\appendix
\onecolumngrid

\renewcommand{\appendixname}{Supplementary Note}
\renewcommand{\thesection}{\arabic{section}}

\setcounter{figure}{0}
\renewcommand{\thefigure}{S\arabic{figure}}
\renewcommand{\theHfigure}{S\arabic{figure}}
\newpage

\section*{Supplementary Material: \\``Non-Convex Optimization via Quantum Tunneling Search on a Photonic Chip"}

\section{Theoretical Analysis}\label{app:theory}
\textbf{Quantum speedups.} {\red{For the family of double-well landscapes defined below ($M_1\ll M_2$), Ref.~\cite{Liu_2023} proves an exponential separation in query complexity between QTW and classical local-query algorithms in its oracle model. The quantum upper bound assumes access to the local ground state of the initial well. Figure~2 of the main text shows a two-dimensional example.}
\begin{equation}\label{eq:double}
f(\mathbf{x})=\left\{\begin{array}{l}
\frac{1}{2} \omega^2\|\mathbf{x}\|^2, \mathbf{x} \in W_{-}, \\
\frac{1}{2} \omega^2\|\mathbf{x}-2 b \mathbf{v}\|^2, \mathbf{x} \in W_{+}, \\
M_1, \mathbf{x} \in B_{\mathbf{v}}, \\
M_2, \text { otherwise.}
\end{array}\right. \tag{S1}
\end{equation}

\begin{problem}[Problem 4.1 of~\cite{Liu_2023}]\label{Problem:1}
For the function $f$ defined in \eqref{eq:double}, given that we only know $0$ is a global minimum, find any point in $W_+$.
\end{problem}

\begin{theorem}[Classical lower bound, Proposition 4.1 of~\cite{Liu_2023}]
    Any classical algorithm will fail, with high probability,  to solve \prob{1} given only $o(e^{\frac{d w^2}{R^2}})$ local queries with or without being restricted in $\mathbb{B}(\mathbf{0}, R)$.
\end{theorem}

\begin{theorem}[Quantum upper bound, Proposition 4.2 of~\cite{Liu_2023}]
    For any dimension $d$, we can always choose appropriate $h, \omega, a, b, H_1, H_2$, and $w$ satisfying previous restrictions, such that, given the local ground state associated with $ W_{-} $ under the choice of $h$ as the initial state, QTW can solve \prob{1} with high probability $1-(1-C)^n$ using only $n O(\operatorname{poly}(d))$ queries, where $0<C<1$ is a constant independent of $d$.
\end{theorem}
}

For a general landscape $f(x)$ consisting of several potential wells, Section 4.2 of~\cite{Liu_2023} shows that the dominant terms affecting the mixing time of SGD and QTW are respectively $\exp(f_{\text{saddle}}-f_{\text{min}})$ and $\exp(\int \sqrt{f(x)}\dd x)$. When the barriers in the landscape are high but thin, the latter term is substantially smaller than the former, \red{so that barrier crossing can favor QTW in the theoretical setting.}

\vspace{.5em}
\textbf{Ensemble average.} Statistical physics tells us that when a system satisfies the conditions for ergodicity, its average result should be equal to the ensemble average. Here, we map the picture of the optimization problem as the Hamiltonian of a system, and the quantum state after local annealing is used as the injection state of the ensemble, and a large number of evolution results of different propagation lengths are averaged.

\vspace{.5em}
\textbf{Quantum annealing speed.} The annealing time scales inversely with the minimal squared spectral gap $\Delta$ of $H(t)$ over all $t\in [0, T]$, typically $t= \mathcal{O} ( 1/\Delta^2)$, with acceleration $t= \mathcal{O} ( 1/\Delta)$~\cite{isakov2016understanding}. Behaving as a spin glass in statistical physics, the energy gap closes drastically with system size $N$ during annealing, which brings exponentially long run time if one aims to precisely follow the adiabatic path. Hence, we are limited to at most polynomial speedups. Besides, it has the defect due to system noise, which is susceptible to long annealing time as well, resulting in the failure to converge to the ground state. Yet we find that the local annealing method can effectively alleviate this problem by annealing in a potential well under the premise that part of the whole picture is known (which is often possible), and then the subsequent search is carried out by quantum tunneling walk.

\vspace{.5em}
\textbf{Initial state significance.} {We now explain why the initial state for quantum tunneling walk should be in a low-energy subspace. The complexity associated with \prob{1} is related to solving a high-dimensional unstructured search problem. Specifically, it can be proven that efficiently solving \prob{1} in $\mathbb{R}^d$ would imply the existence of a quantum algorithm capable of solving an unstructured search problem whose size $N$ is exponential in $d$ using a number of queries polynomial in $d$. Formally speaking, if a quantum algorithm solves \prob{1} with a polynomial query complexity in $d$, then it could solve an unstructured search of size $N = 2^{\Omega(d)}$ within $O(\mathrm{poly}(\log N))$ queries.

However, this result contradicts the established lower bound by Bennett et al.~\cite{Bennett_1997}, which states that the quantum query complexity for an unstructured search of size $N$ is at least $\Omega(\sqrt{N})$. Consequently, we have

\begin{theorem}[Proposition 4.3 of~\cite{Liu_2023}]
    No quantum algorithm can solve \prob{1} using query complexity polynomial in $d$ without priori.
\end{theorem}

Thus, to solve \prob{1} we require a local ground state as the initial near-optimal input.
}

\vspace{.5em}
\textbf{Stochastic gradient descent.} The pseudo-code for our stochastic gradient descent algorithm in a discrete potential landscape is presented in \algo{SGD}. After smoothing the potential wells with a gap of 0.1. We set up 1000 groups of experiments, each running 800 steps. Then we empirically set the learning rate $s=3$ for fair comparison with the quantum learning rate $h=1$ with near risk/distance accuracy (see Standard 4.1 and 4.2 of~\cite{Liu_2023} for detailed criteria). Note that a too-small learning rate value may result in failure in tunneling. \red{The initial position of each trajectory is randomly sampled from a discrete probability distribution representing the spatial profile of the locally annealed QTS input state. The weights used in \algo{SGD} form a triangular profile on $[0,4]$ centered at $x=2$}. \red{Normal noise is added only if} the magnitude of gradient is smaller than $10^{-3}$. When moving outside $[0,28]$, we implement a rebound. For every position, we only round to one decimal place to adapt to the discrete landscape.

\begin{algorithm}[!htbp]
\caption{Stochastic gradient descent algorithm in a discrete potential landscape}
\label{algo:SGD}
\begin{algorithmic}[1]
    \REQUIRE $\text{smoothed\_potential\_landscape with gap}=0.1$
    $\text{num\_simulations} = 1000$, $\text{num\_steps} = 800$, $\text{learning\_rate} = 3$, 
    \STATE $\text{positions} \leftarrow [~]$
    \STATE $\text{x} \leftarrow \text{linspace}(0, 4, 1000)$
    \STATE $\text{weights} \leftarrow \text{where}(x, x \leq 2; 4 - x, x>2)$
    \STATE $\text{weights} \leftarrow \text{weights} / \text{sum(weights)}$
    \FOR{$i = 1$ to $\text{num\_simulations}$}
        \STATE $\text{position} \leftarrow \text{random.choice}(x, \text{p}=\text{weights})$
        \FOR{$j = 1$ to $\text{num\_steps}$}
            \STATE $\text{gradient} \leftarrow \text{increment}[\text{index of closest point}]$
            \IF{$|\text{gradient}| < 10^{-3}$}
                \STATE $\text{noise} \leftarrow \text{random.normal}()$
            \ELSE
                \STATE $\text{noise} \leftarrow 0$
            \ENDIF
            \STATE $\text{position} \leftarrow \text{position} - \text{learning\_rate} * (\text{gradient} + \text{noise})$
            \IF{$\text{position} < 0$}
                \STATE $\text{position} \leftarrow -\text{position}$
            \ELSIF{$\text{position} > 28$}
                \STATE $\text{position} \leftarrow 56 - \text{position}$
            \ENDIF
        \ENDFOR
        \STATE $\text{positions.append(round(position, 1))}$
    \ENDFOR
    \STATE count positions distribution
    \ENSURE {unique\_positions}, {probability\_distribution}
\end{algorithmic}
\end{algorithm}

\vspace{.5em}
\textbf{\red{Benchmark accounting and scope.}}
\red{The SGD traces shown in the main text are sampled software-trajectory distributions, whereas the photonic curves are measured output-intensity fractions. Their horizontal axes are fundamentally different: one counts iterative updates and the other reports fabricated propagation length. The plotting factor of 60 is a display convention and not an independently established comparison of resources. For a matched-budget comparison, one must separately report the number of calls to the discrete objective function or its gradient, all hyperparameter-search evaluations, the number of trajectories and random seeds, the quality metric at each budget, and the measured execution time. For a physical comparison, optical input/output handling and the one-time costs of sampling the objective, calibrating and writing the waveguide arrays must also be included. The data presented here do not allow such an end-to-end speedup estimate.}
\red{In \fig[b]{4}, the root-mean-square position $\sqrt{\sum_j x_j^2 P_j(z)}$ measures the extent of the intensity distribution relative to the coordinate origin, near the injection well. Intensity at any site with large $x_j$ increases this second moment, even when little intensity reaches a target well.}

\section{Experimental Details}\label{app:details}

\textbf{\red{Optical excitation, CCD imaging, and computer acquisition.}} \red{In the QTS experiments, the incident laser beam is directly focused into a selected input waveguide through a 20$\times$ microscope objective. The optical measurements were carried out at $21\,^\circ\mathrm{C}$ with a nominal input optical power of \SI{1}{mW}. The quoted input power is not assumed to equal the independently unmeasured power coupled into the waveguide. The spatial optical-intensity distribution emerging from the output facet is recorded with a charge-coupled device (CCD) camera, and the CCD images are acquired by a computer for subsequent extraction of waveguide-resolved intensities and calculation of the normalized output distributions. The computer is used for data acquisition and post-processing; the waveguide propagation-constant profile remains fixed by fabrication rather than being electronically reprogrammed during measurement. The normalized distributions represent classical-light intensity fractions, not probabilities extracted from single-photon counting. The two-photon coincidence maps discussed in \app{open} are numerical simulations and do not describe detectors or measurements used for the main-text experiments.}

\vspace{.5em}
\textbf{\red{Propagation-length sampling and interpretation.}} \red{The coordinate $z$ in the Hamiltonian propagation model is a spatial coordinate along the chip. Experimental evolution snapshots are output measurements taken from arrays fabricated for the respective propagation lengths, rather than time-resolved movies from a single output camera. For the two-well data in Fig.~2, the length series spans \SIrange{5}{90}{mm} in \SI{5}{mm} increments; the averaged distribution is the arithmetic mean across these measured samples. The varying-$H$ curve in \fig[a]{4} shows the arithmetic mean of the last eight recorded intensity distributions. The measurement end length is set by the available structures and modulation range, not by an online certificate of optimality; determining a stopping rule for unknown landscapes remains an open issue.}
\red{In \fig[c]{2}, intensity is visible in the right well at the sampled length $z=\SI{55}{mm}$. Assigning a first detectable arrival length would require a specified intensity threshold and would be limited by the spacing between sampled lengths. The running averages of well occupation in \fig[b]{3} describe redistribution over the measured length series. Estimating a mixing length would additionally require a reference distribution, a distance measure and a convergence tolerance.}

\vspace{.5em}
\textbf{\red{CCD pixel extraction, intensity normalization, and quantitative metrics.}} \red{For the one-dimensional 29-site waveguide dataset, the CCD output is processed as a pixel-intensity matrix. The supplied MATLAB processing workflow estimates a common background level from image corners and subtracts it from the image, with negative corrected pixel values clipped to zero. For position calibration, the image is first averaged in small pixel blocks; sufficiently bright local maxima provide initial waveguide anchors. Their positions are refined locally and used to calibrate the regular 29-site array, so that weaker channels are included at the predicted lattice positions rather than being discarded simply because their peaks are faint. A rectangular region of interest (ROI) is assigned to each waveguide, and its background-corrected pixel values are summed after a locally defined intensity threshold is applied. The extracted site intensities are then normalized over all 29 waveguides.}
\red{Let $D_{m}(u,v)$ be the raw pixel value at propagation length $z_m$, $B_m$ the estimated frame background and $\widetilde D_m(u,v)=\max[D_m(u,v)-B_m,0]$. With a calibrated rectangular site ROI $\mathcal R_j$ and site-specific threshold $\tau_{j,m}$, the processed intensity is}
\red{
\begin{equation}
I_j(z_m)=\sum_{(u,v)\in\mathcal R_j}\widetilde D_m(u,v)
\,\mathbf{1}\!\left[\widetilde D_m(u,v)\geq\tau_{j,m}\right].
\label{eq:exp-roi-integration}
\end{equation}
}
\red{The normalized optical intensity distribution and its arithmetic mean over $M$ measured propagation lengths are}
\red{
\begin{equation}
P_j(z_m)=\frac{I_j(z_m)}{\sum_k I_k(z_m)},\qquad
\overline P_j=\frac{1}{M}\sum_{m=1}^M P_j(z_m).
\label{eq:exp-normalization}
\end{equation}
}
\red{Here $P_j$ is an experimentally inferred classical optical-intensity fraction, not a single-photon detection frequency. Its accuracy depends on background estimation, channel calibration, pixel thresholding, potential overlap between neighboring ROIs, and CCD fluctuations.}

\red{For a well represented by a predefined set $W$ of waveguide indices, the well occupation and its running length-average are}
\red{
\begin{equation}
Q_W(z_m)=\sum_{j\in W}P_j(z_m),\qquad
A_W(z_n)=\frac{1}{n}\sum_{m=1}^n Q_W(z_m).
\label{eq:exp-well}
\end{equation}
}
\red{A possible localization-contrast measure, with $\bar P_W^{\mathrm{site}}=|W|^{-1}\sum_{j\in W}\overline P_j$ and $\bar P_{\neg W}^{\mathrm{site}}=|W^c|^{-1}\sum_{j\notin W}\overline P_j$, is $\mathcal C_W=(\bar P_W^{\mathrm{site}}-\bar P_{\neg W}^{\mathrm{site}})/(\bar P_W^{\mathrm{site}}+\bar P_{\neg W}^{\mathrm{site}})$. For the global-minimum example in Fig.~5, the single-site observable is $P_{14}(z)$. These measures quantify the share and contrast of optical intensity in regions that were identified from the prescribed test landscape; without repeated, independent runs of an explicitly defined retrieval rule, they should not be called empirical algorithmic success rates.}

\vspace{.5em}
\textbf{\red{Representative signal-to-noise ratio of the CCD images.}}
\red{The optical output patterns at successive propagation lengths were captured sequentially by the CCD camera; the resulting image series are already displayed in the supplementary figures (e.g., Figs.~\ref{fig:A3} and \ref{fig:A5}--\ref{fig:A8}). We estimate a typical background-limited spatial signal-to-noise contrast from one representative raw image of the $29\times1$ waveguide array. After cropping the original $1448\times1928$ pixels to $1440\times1920$ pixels following our image-processing procedure, four corner patches, each of $200\times200$ pixels, were used to estimate the background mean $\mu_{\mathrm{bg}}$ and spatial standard deviation $\sigma_{\mathrm{bg}}$ before any background subtraction or thresholding. For a resolved waveguide spot $j$, the mean raw intensity $\langle D\rangle_{\mathcal S_j}$ was determined in a $9\times9$-pixel region around its center. We define the background-limited spatial signal-to-noise ratio (equivalently, a signal-to-background contrast-to-noise measure) as}
\red{
\begin{equation}
\mathrm{SNR}^{\mathrm{spatial}}_{j,\mathrm{dB}}
=20\log_{10}\!\left(\frac{\langle D\rangle_{\mathcal S_j}-\mu_{\mathrm{bg}}}{\sigma_{\mathrm{bg}}}\right).
\label{eq:exp-spatial-snr}
\end{equation}
}
\red{In the representative frame, $\mu_{\mathrm{bg}}=56.93$ and $\sigma_{\mathrm{bg}}=3.92$ CCD digital counts (DN). The median value across 11 clearly resolved waveguide spots was $49.6\,\mathrm{dB}$, giving a typical spatial signal-to-noise contrast of approximately $50\,\mathrm{dB}$ for this example. The $9\times9$-pixel region is used only to characterize the signal-to-background contrast; the output distributions plotted in the manuscript are extracted using the background-corrected, thresholded integration of all 29 waveguides described above. As the numerical estimate is based on one representative frame rather than a separate analysis of every supplementary image, it should not be interpreted as a measured time-domain SNR or as a pooled value for all propagation lengths.}

\vspace{.5em}
\textbf{\red{Fabrication overhead and scope of scalability.}} \red{Fabricating a complete chip in the present experiments required approximately 5--10 hours. This is the overall chip fabrication time, including fabrication operations beyond a single waveguide scan; it is not the propagation time or the time required for a single 10-cm waveguide. Both the relative propagation constants and their longitudinal variations are encoded during laser writing; a different sampled objective generally needs a different fabricated array. For a $d$-dimensional grid with $n_k$ samples on each axis, the site count grows as $N=\prod_{k=1}^d n_k$. The experimental hardware therefore offers a fixed analog propagation process for a given landscape, rather than reconfigurable evaluation of an arbitrary unknown function. A meaningful end-to-end cost estimate must include objective sampling and layout design, calibration structures, writing each array, optical alignment, acquisition and digital analysis, as well as the number of structures produced per substrate. We do not infer these costs from propagation length alone.}

\vspace{.5em}
\textbf{\red{Evidence and controls for tunneling-inspired transport.}} \red{Coherent coupling also transports light through an unstructured waveguide array. Consequently, output intensity at a distant minimum is consistent with the engineered Hamiltonian dynamics but, on its own, does not isolate the contribution of a programmed potential barrier. A discriminating control would compare the same initial field, lattice geometry, coupling coefficient and propagation lengths for a structured barrier, a barrier-reduced or barrier-free profile, and a uniform array, evaluating $Q_W$ and the temporal length dependence under comparable conditions. No additional barrier-control data are inferred from the measurements currently reported.}

\textbf{Waveguide setup.} \red{The coupled-mode equations for the coherent field amplitudes in the waveguides have the same form as the evolution equations for the site amplitudes $c_j(z)$ in the single-excitation sector of the tight-binding Hamiltonian. Thus, the normalized output intensity at site $j$ measures the optical counterpart of $|c_j(z)|^2$. At the grid point $x_j$, the design satisfies $\beta_j(z)-\beta_{\mathrm{ref}}(z)=s(z)f(x_j)$, where $s(z)>0$ sets the potential scale and the common offset $\beta_{\mathrm{ref}}(z)$ contributes only a global phase.}
We can freely set up different 3D structures beyond uniform waveguide arrays. For the landscape $f(x)$ in the homogeneous space solved in this paper, the coupling coefficients are set to be the same, but it is conceivable that varying values of $C_{i, j}$ open up the possibility of solving the function in more complex spaces, such as mapping distorted waveguide arrays into manifolds or functional spaces with special topologies~\cite{wang2022topologically}.

\begin{figure}[!htbp]
    \centering
    \includegraphics[width=0.5\linewidth]{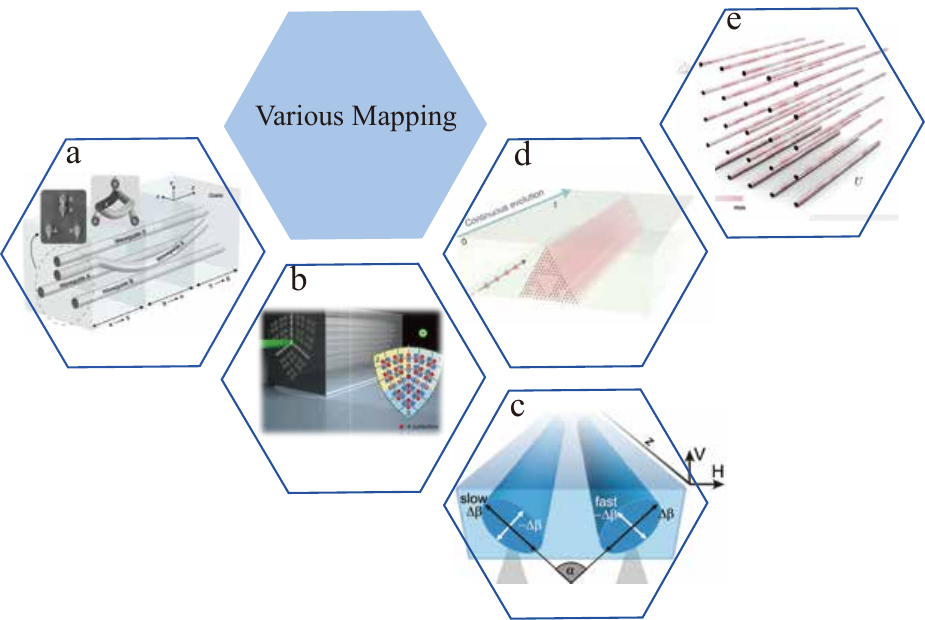}
    \caption{{Various approaches of waveguide Hamiltonian mapping. \textbf{a} to \textbf{e} respectively represent the Hamiltonian mapping in a topological picture~\cite{Zhang_2022}, the orbital angular momentum (OAM)-encoded space~\cite{hu2024topological}, dimensional enhancement through polarization encoding~\cite{doi:10.1126/science.ado8192}, the scenario under fractal structures~\cite{xu2021quantum}, and the Haar-random space~\cite{Tang_2022}.}}
    \label{fig:mapping}
\end{figure}

\vspace{.5em}
\textbf{Fabrication and modulation.}
\red{The shared femtosecond-laser writing platform uses a Yb: KGW regenerative-amplifier source delivering \SI{515}{nm} pulses of \SI{290}{fs} duration at a repetition rate of \SI{1}{MHz}. A cylindrical lens of focal length \SI{700}{mm} is used for beam shaping before focusing through a 50$\times$ objective (NA = 0.55) into glass mounted on a computer-controlled high-precision translation stage.}
\red{The waveguide arrays are fabricated in Corning Eagle XG glass using femtosecond laser writing. The nominal individual waveguide path length is approximately \SI{10}{cm}, and the typical laser writing speed is approximately \SI{1}{cm/s}. Since the target on-site propagation constants are encoded through writing-speed modulation, the local scan speed is intentionally varied from this typical value as required by each programmed profile. Before writing the objective-dependent profiles, the nearest-neighbor coupling coefficient $C$ is calibrated with matched two-waveguide structures. In the ideal coupled-mode model, the normalized intensities oscillate as $I_a(z)=\cos^2(Cz)$ and $I_b(z)=\sin^2(Cz)$, so their normalized difference is} 

\begin{equation}
    \frac{I{(a)}-I{(b)}}{I{(a)}+I{(b)}}=\frac{\cos^{2}(Cz)-\sin^{2}(Cz)}{\cos^{2}(Cz)+\sin^{2}(Cz)}=\cos(2Cz).
    \label{coupling coefficient} \tag{S2}
\end{equation}

\begin{figure*}
    \includegraphics[scale=0.17]{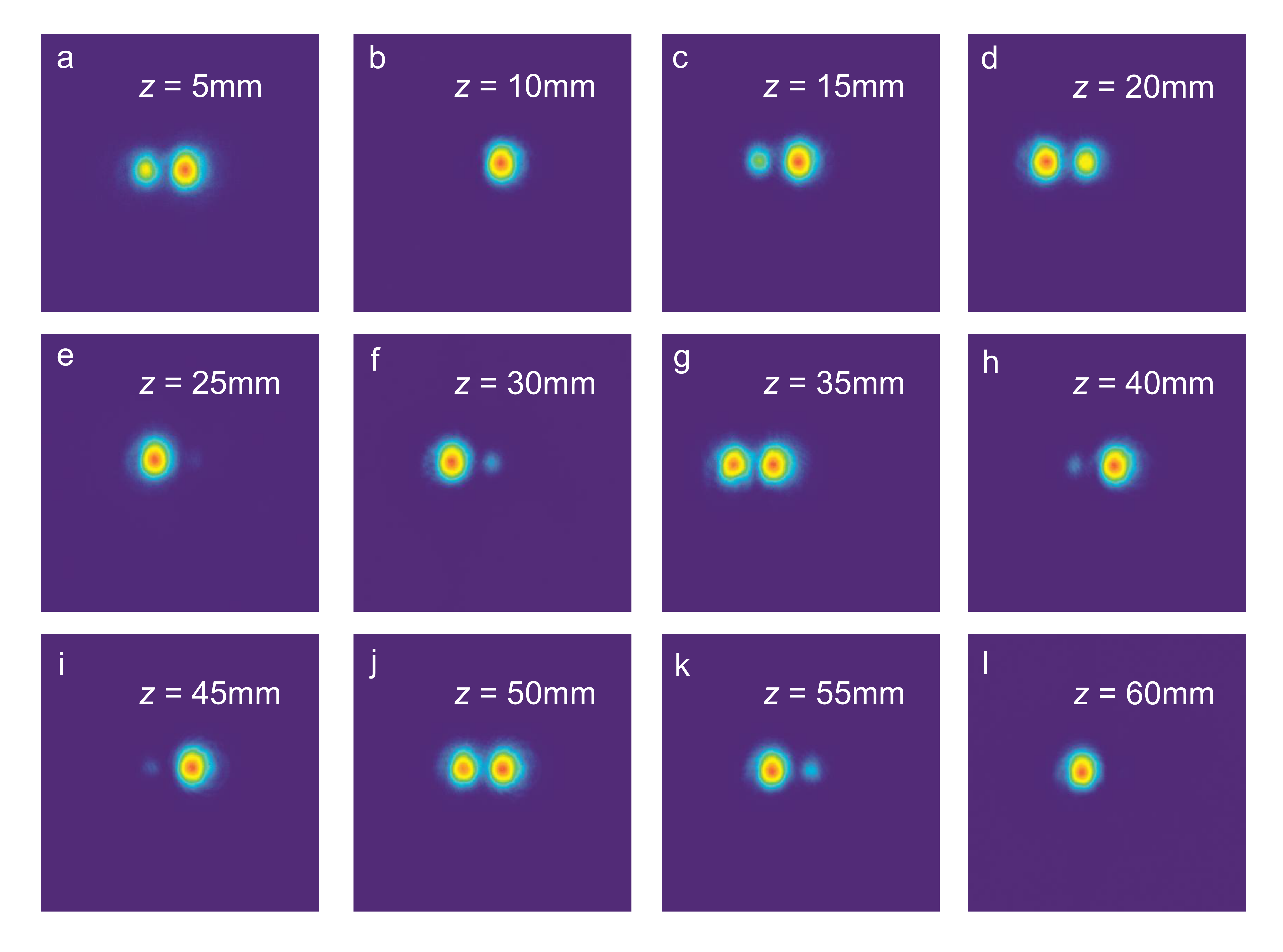}
    \caption{Coupling coefficient calibration. Experimentally measured intensity of two coupling waveguides with coupling length ranging between $\SIrange[range-phrase={-}]{5}{60}{mm}$. The period of intensity oscillation is utilized to calibrate the coupling coefficient according to \eqref{coupling coefficient}.}
    \label{fig:A1}
\end{figure*}

\begin{figure*}
    \includegraphics[scale=0.45]{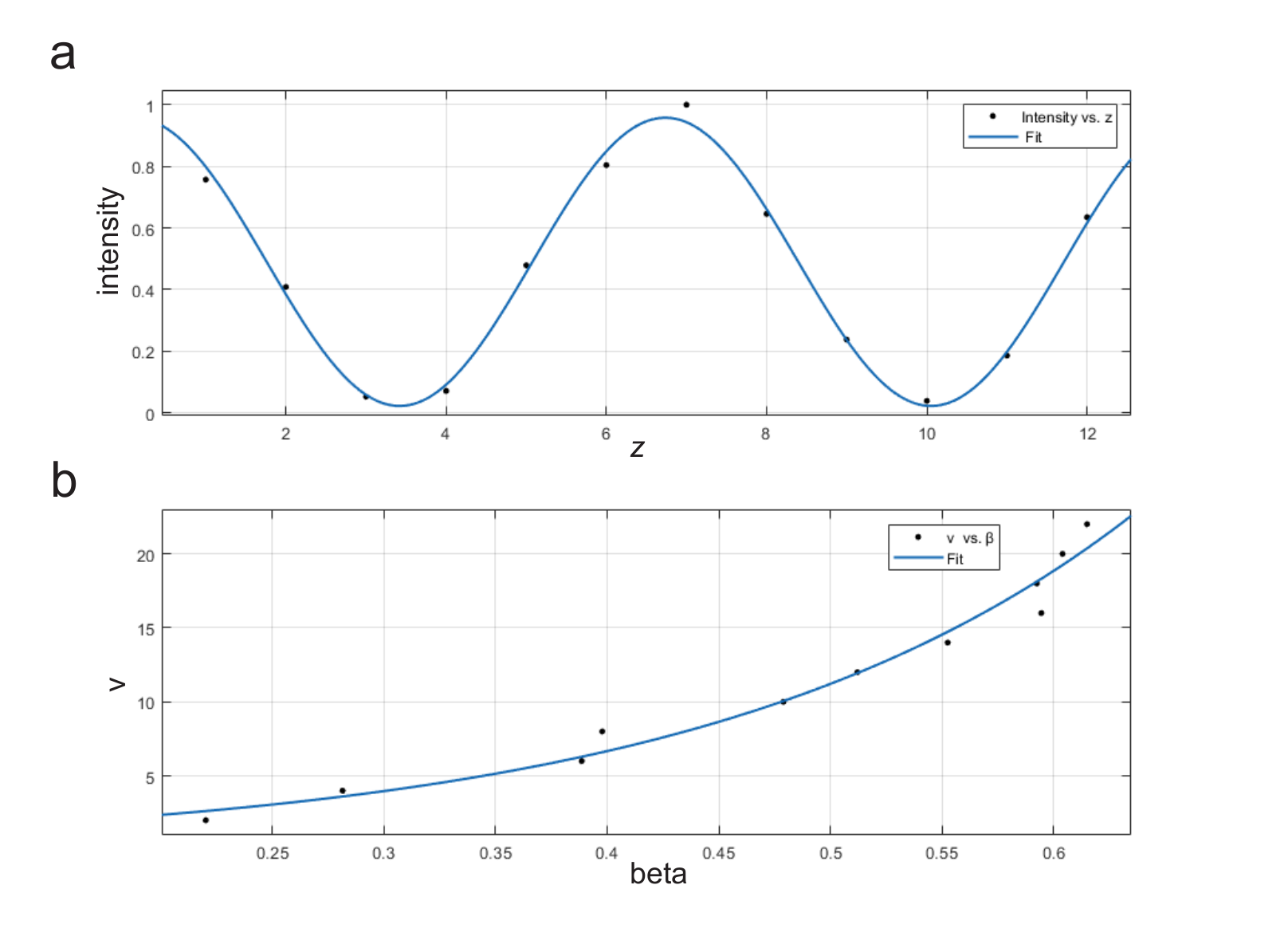}
    \caption{Coupling coefficient and $\beta$ - speed fitting. \textbf{a} The intensity oscillation between two coupled waveguides, and the interval is set to $\SI{1}{mm}$, which yields the coupling coefficient $C=0.47$. \textbf{b} The results of the exponential fit for writing and $\beta$. Fitting result $v = 0.8788e^{5.1233\cdot \beta}$.}
    \label{fig:A2}
\end{figure*}

\red{A series of calibration pairs with different coupling lengths was fabricated, and the output of each guide was recorded by CCD imaging. Fitting the intensity oscillation yields $C$; the measured oscillation curves are shown in \fig{A1} and \fig{A2}. The nominal coupling coefficient used in the one-dimensional simulations and several fabricated arrays is approximately \SI{0.47}{mm^{-1}} (see \fig{A2}).}

\red{The on-site propagation-constant detuning is programmed by changing the laser writing speed. Compared with writing-power modulation, this was the chosen implementation for the arrays studied here. To calibrate the speed dependence, one guide of a two-guide coupler is fabricated at a varied writing speed, and the output intensity oscillation is recorded. Denoting the propagation-constant difference between the two guides by $\Delta\beta=\beta_a-\beta_b$, coupled-mode theory gives}
\red{
\begin{equation}
    C_\mathrm{eff}=\sqrt{(\Delta\beta/2)^2+C_0^2}, \tag{S3}
\end{equation}
}
\red{where $C_0$ is the matched-guide coupling coefficient. Thus $|\Delta\beta|$ follows from the independently fitted values of $C_{\mathrm{eff}}$ and $C_0$; the choice of sign must be fixed by the writing-speed calibration. Figure~\ref{fig:A2} reports an empirical relation between writing speed and the programmed detuning. The same calibration can be used to prescribe spatially varying $\Delta\beta_j(z)$ by varying the writing speed as a function of $z$. The calibration fit is empirical, so model mismatch and writing deviations can perturb the implemented Hamiltonian. These variations are fixed in the fabricated glass and cannot be updated during an experimental run.}

\vspace{.5em}
\textbf{Three-well landscape.} \red{For the three-well landscape, length-resolved experimental output images are given in \fig{A3}; their spatial probability summaries and occupation of the rightmost well are compared with simulation in \fig{A4}.}

As shown on the left side of \fig{A4}, we present the probability distribution of the system after $z=\SI{90}{mm}$ evolution for the three-well landscape. The right side of \fig{A4} shows the simulated and experimental probability accumulating in the rightmost potential well. The probability accumulating in the rightmost well is also consistent with the simulation result.

\red{The uncertainty bands shown in the simulations in \fig{A4} are obtained by perturbing the programmed propagation-constant parameter with zero-mean Gaussian noise, as described in that figure caption. This is a numerical sensitivity test and should not be interpreted as a measured fabrication tolerance, detector SNR or confidence interval for repeated chips. An experimental fabrication tolerance would require repeated propagation-constant or writing-speed calibrations, preferably across nominally identical structures. Nor does classical coherent illumination imply that output imaging is free from detector noise, source fluctuations, or alignment drift.}

Similarly, we present the experimental snapshots of quantum tunneling search in a five-well landscape in \fig{A5}, in a 2D double-well landscape in \fig{A6}, in a random landscape with constant and varying Hamiltonian respectively in \fig{A7} and \fig{A8}.

\vspace{.5em}
\textbf{Random landscape.} \red{The available refractive-index modulation limits the increase in the programmed potential at the final stage. To characterize this stage, we averaged the last eight measured intensity distributions for the varying-$H$ curve in \fig[a]{4}.}

\red{Varying the ratio of on-site detuning to coupling along $z$ changes the instantaneous modes of the array and can redistribute intensity among wells. We assess this effect through localization over the available propagation lengths. Increasing the ratio in a lossless, coherently evolving array does not by itself cause relaxation to the instantaneous ground state.}

\vspace{.5em}
\textbf{Exotic landscape.} After performing quantum tunneling search in various multi-well landscapes, we test its performance under some typical exotic non-convex functions, as shown in \fig{A11}. Here we also adopt the local annealing with quantum tunneling walk scheme, with the range of all the landscapes $f(x)$ restricted to $[-1,1]$ and the coupling coefficient $ C$ set to 0.4.

\section{Future Directions}\label{app:open}
\red{\textbf{Dimension dependence.} In our waveguide arrays, the transverse site coordinates encode the optimization variables, while the longitudinal coordinate sets the propagation length. This mapping gives the one- and two-dimensional landscapes studied here. Extending the same grid encoding to $d$ variables with $n$ samples per variable would require $n^d$ waveguides. An open question is whether additive structure or low effective dimension can reduce the number of optical modes needed to represent a landscape while preserving tunneling between its wells.}

\vspace{.5em}
\red{\textbf{Query cost of landscape encoding.} To fabricate the arrays used here, we first evaluate the objective function at each grid point and use those values to set the waveguide propagation constants. For a grid of $N$ sites, this requires $N$ objective evaluations. Once these values are available, a single scan identifies the minimum on the grid. For objectives with a known functional form, a direction for future work is to determine whether the optical potential can be generated directly from the parameters of the function, without first evaluating it on the full grid.}

\vspace{.5em}
\textbf{Photon enhancement.} \red{Beyond the coherent-light implementation demonstrated in the main text, multi-photon states provide a possible route for investigating quantum contributions to quantum tunneling search. We therefore analyze the effects of photon indistinguishability and coincidence-counting readout within the same Hamiltonian framework. The simulated coincidence distributions in \fig{A10} are a prospective extension of the present experiment.}

The effect of different injection states and the advantages of coincidence counting in the photonic regime due to photon indistinguishability are further analyzed below. \red{For each input state in \fig{A9}, we simulate propagation on the same landscape at 60 lengths from \SI{0}{mm} to \SI{295}{mm} in \SI{5}{mm} steps.} As demonstrated in the left side of \fig{A9}, for the direct injection state without local annealing, its energy is too high to converge to the ground state, thus can not give a faithful depiction of the minimum. However, combined with the local annealing state, what we call distinguishable photon injection, the search result renders a more exact result owing to the depression of lower energy state. \red{The running averages of the right-well occupation in \fig[b]{A9} illustrate the sensitivity of the search dynamics to the input-state preparation.}

\red{We next consider simulated coincidence distributions for indistinguishable- and distinguishable-photon inputs.} For both cases, the injection state is the combination of local annealing state and single photon injection at $x=14$. The landscape of evolution is the same as the three-well mentioned before, and the coincidence counting is the average over $z=\SIrange[range-phrase={-}]{0}{195}{mm}$. It can be seen from \fig{A10} that indistinguishable photon search presents a similar performance to the distinguishable photon search in position $(2,14)$, which marks the position of the wells on either side. The wells at $x = 2$ and $x = 14$ have a higher probability of being found for all input states, but the well at the $x = 8$ position becomes a key measure for comparing different injection methods. \red{At representative coordinates such as $(8,8)$, $(2,8)$, and $(14,8)$, the indistinguishable-photon case exhibits higher coincidence weight under the conditions considered here. Only one of the two input components undergoes local annealing, yet the coincidence distribution retains features at the positions of the minima. A systematic assessment of whether such correlations provide a computational or metrological advantage will require dedicated quantum-photonic experiments and comparisons with appropriate classical-wave baselines.}

\newpage

\begin{figure*}
    \includegraphics[scale=0.3]{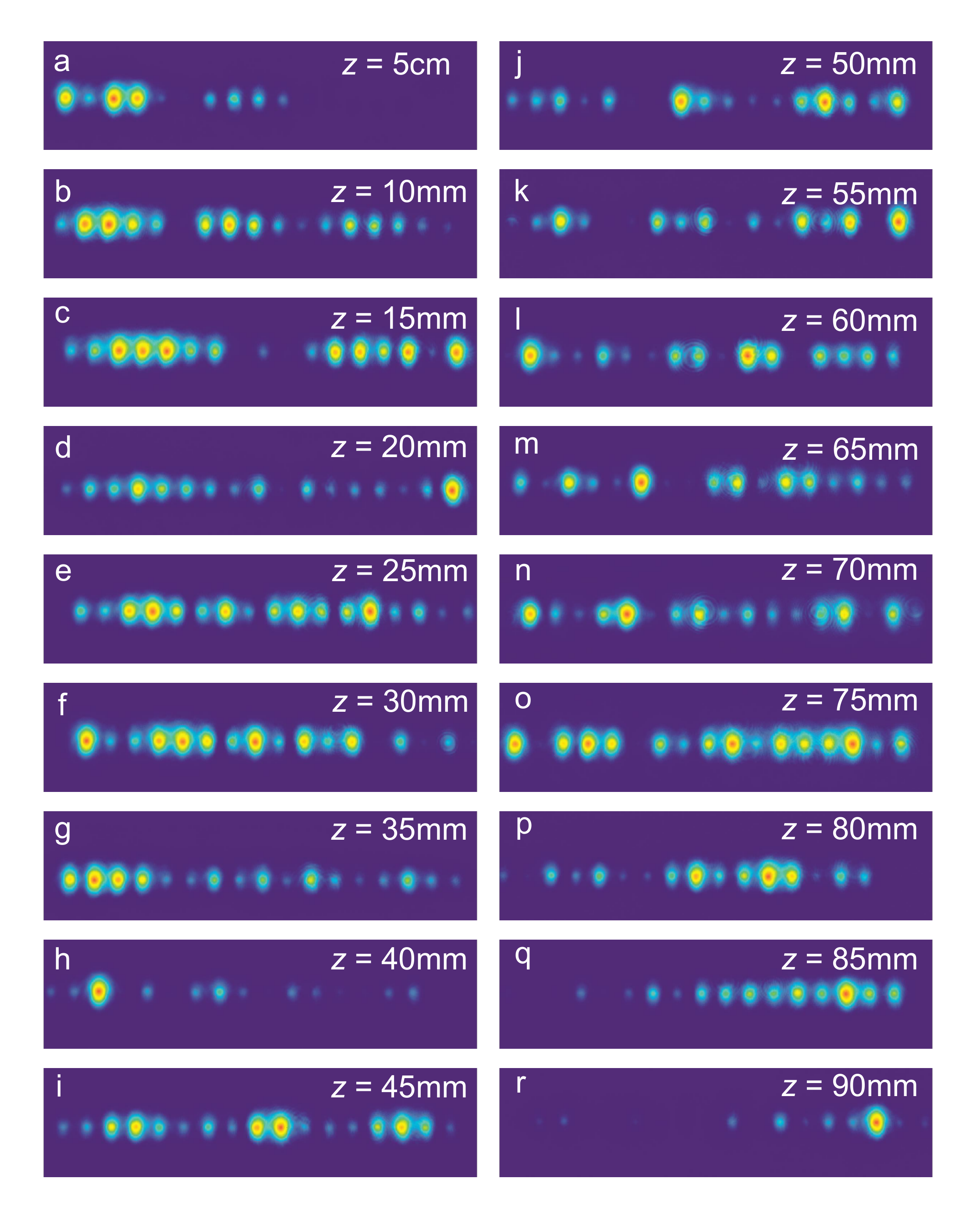}
    \caption{Experimental quantum tunneling search in a three-well landscape. The measurement interval is set to 1 mm, and the total length $\SIrange[range-phrase={-}]{2}{22}{mm}$ is taken for the average.}
    \label{fig:A3}
\end{figure*}

\begin{figure*}\label{fig:three}
    \includegraphics[width=\linewidth]{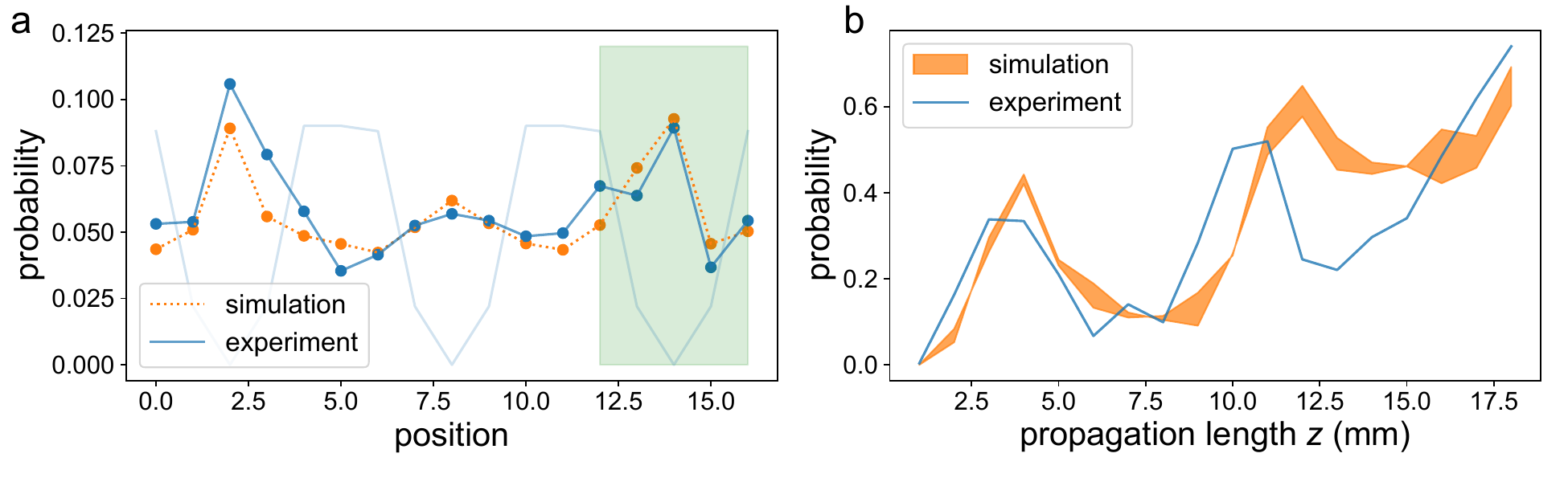}
    \caption{Three-well landscape simulation and experimental results. \textbf{a} The left side presents the average probability distribution. \textbf{b} The right side compares the accumulated probability in the rightmost potential well. \red{The simulation uncertainty bars represent a numerical sensitivity test rather than an independently measured fabrication tolerance. The numerical model adds zero-mean Gaussian perturbations with standard deviation $0.2$ to the programmed $\beta$ parameter.}}
    \label{fig:A4}
\end{figure*}

\begin{figure*}
    \includegraphics[scale=0.3]{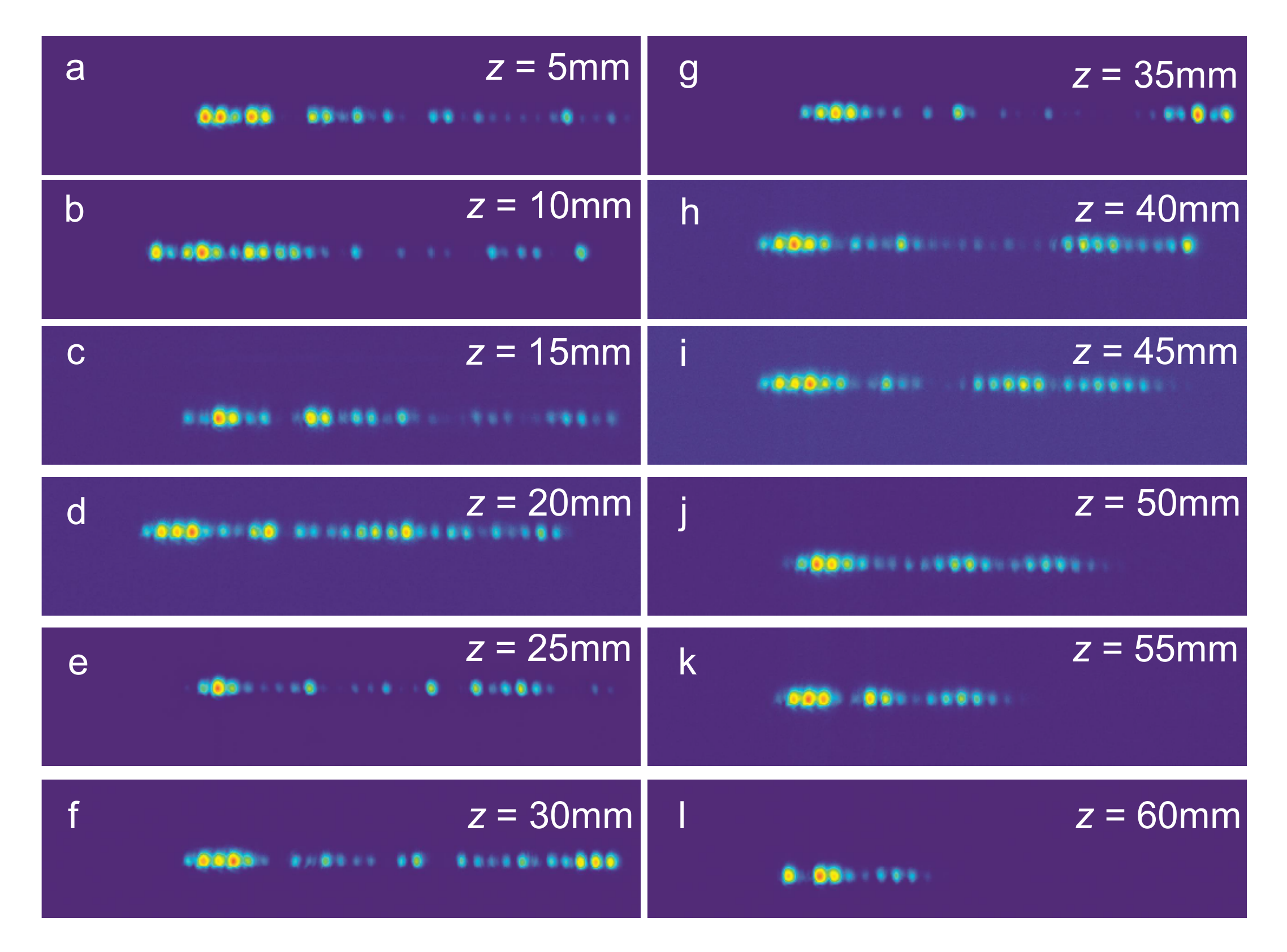}
    \caption{Experimental quantum tunneling search in a five-well landscape. Please refer to Figure 3 in the main text for the landscape.}
    \label{fig:A5}
\end{figure*}

\begin{figure*}
    \includegraphics[scale=0.15]{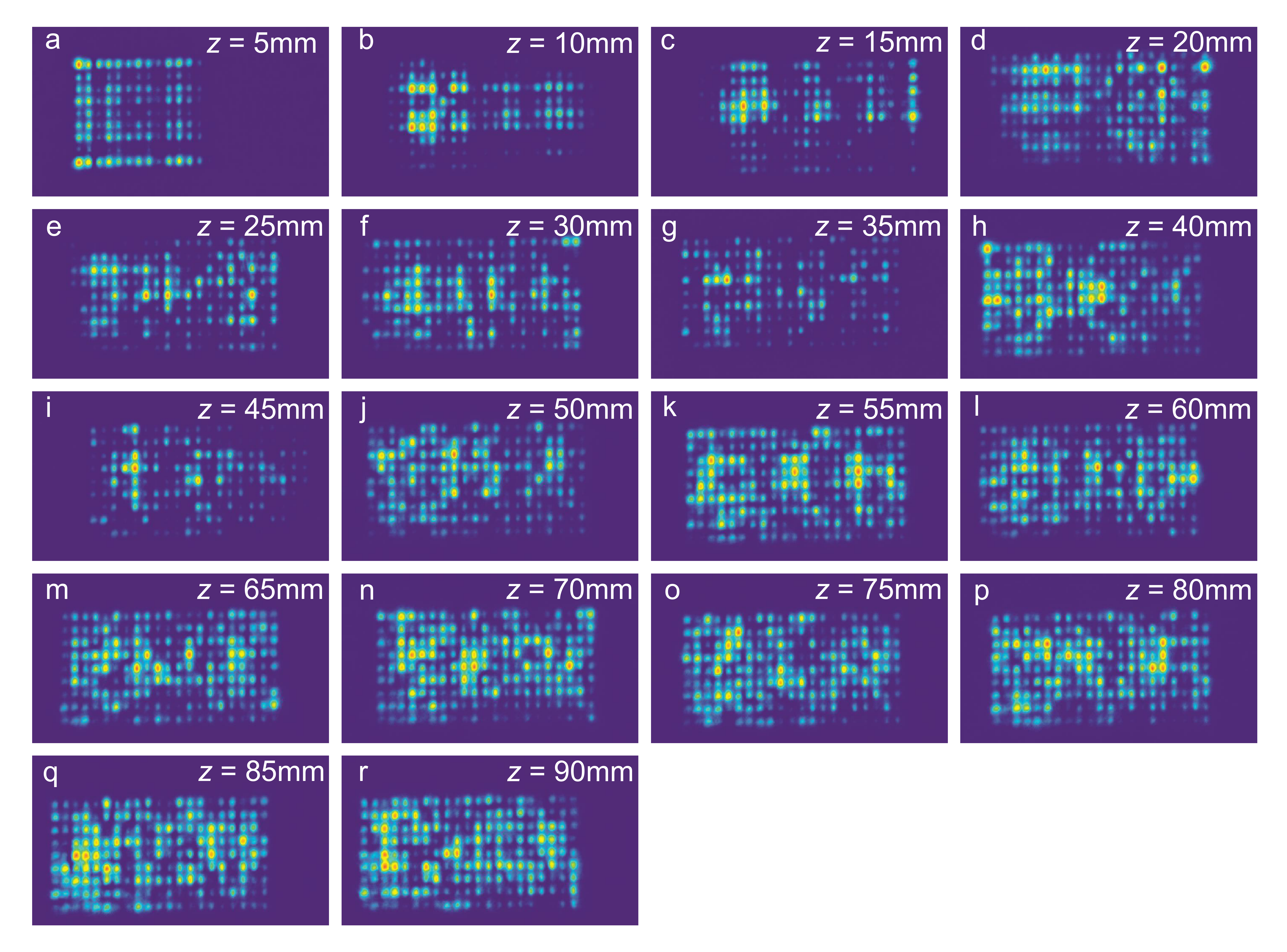}
    \caption{Experimental quantum tunneling search in 2D double-well landscape. $\beta$ modulation ranges from $\SIrange{0}{1}{mm^{-1}}$ and coupling coefficient was set to $\SI{0.47}{mm^{-1}}$. Please refer to Figure 2 in the main text for the landscape.}
    \label{fig:A6}
\end{figure*}

\begin{figure*}
    \includegraphics[scale=0.32]{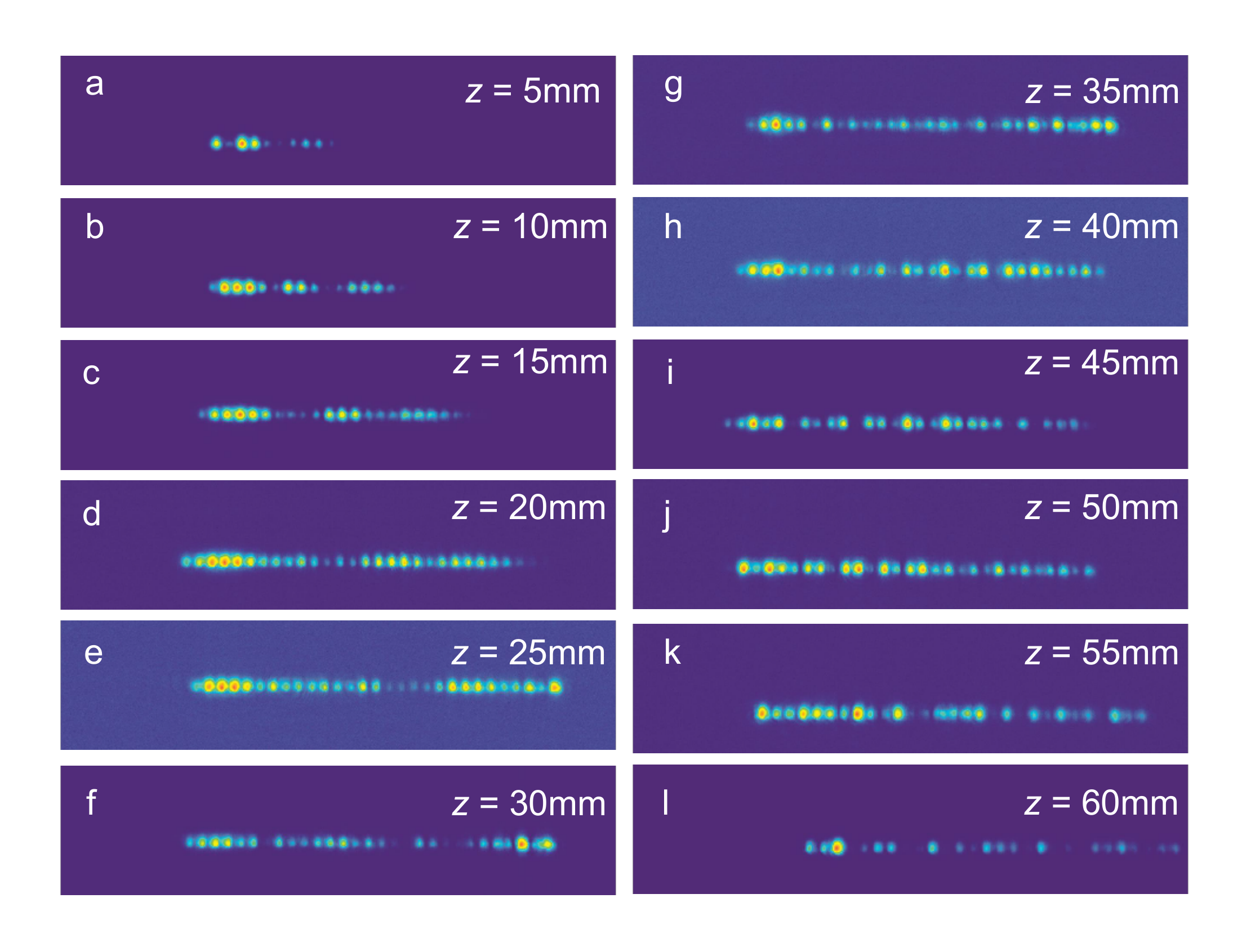}
    \caption{Experimental quantum tunneling search in random landscape search with local annealing and constant Hamiltonian. $\Delta \beta$ modulation ranges from $\SIrange{0}{1}{mm^{-1}}$  and coupling coefficient was set to $\SI{0.47}{mm^{-1}}$. Please refer to Figure 4 in the main text for the landscape.}
    \label{fig:A7}
\end{figure*}

\begin{figure*}
    \includegraphics[scale=0.32]{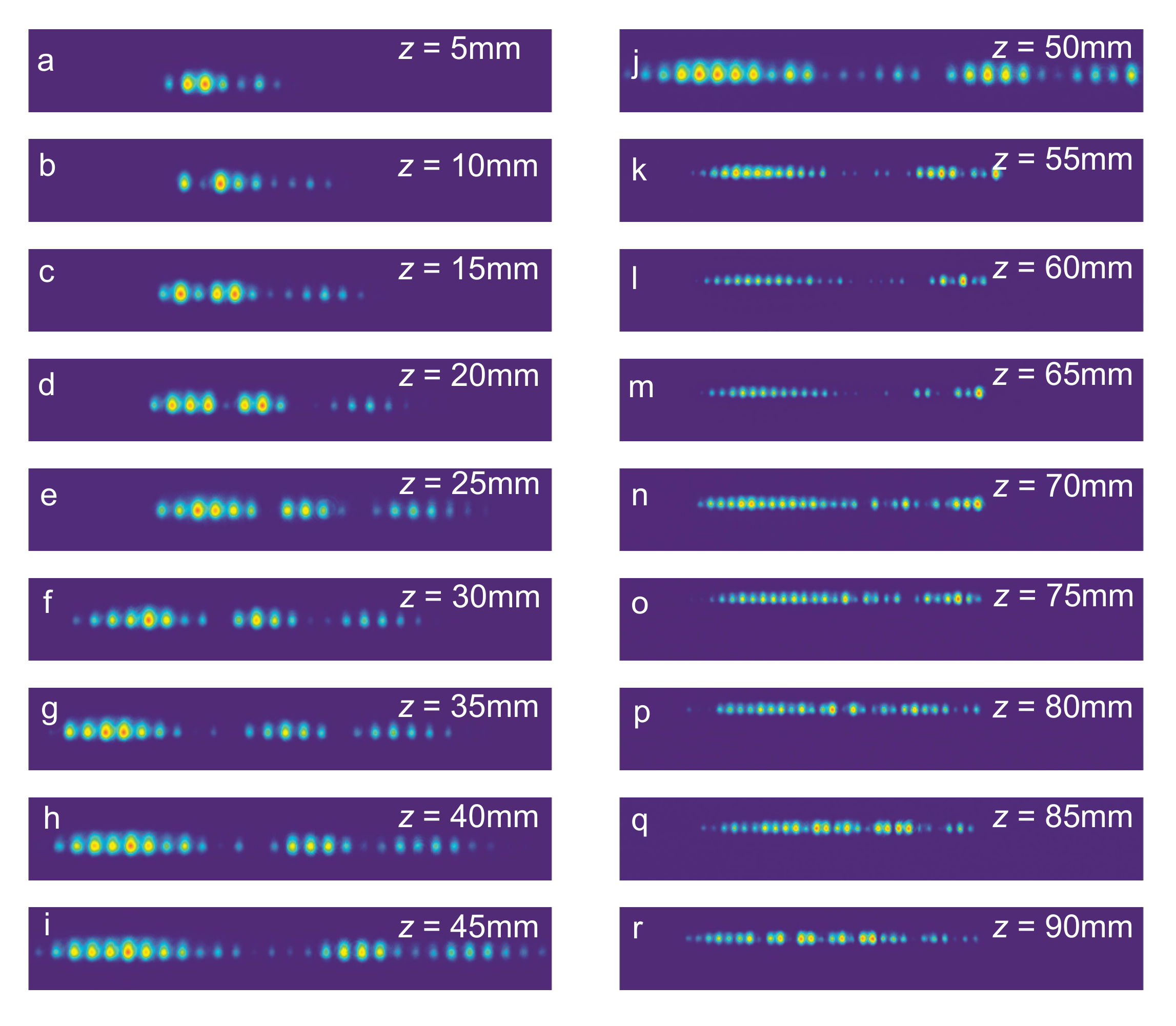}
    \caption{Experimental quantum tunneling search in random landscape search with local annealing and varying Hamiltonian. $\Delta \beta$ modulation ranges from $\SIrange{0}{1}{mm^{-1}}$ and coupling coefficient was set to $\SI{0.47}{mm^{-1}}$. Please refer to Figure 4 in the main text for the landscape.}
    \label{fig:A8}
\end{figure*}

\begin{figure*}[h]
    \centering
    \includegraphics[width=\linewidth]{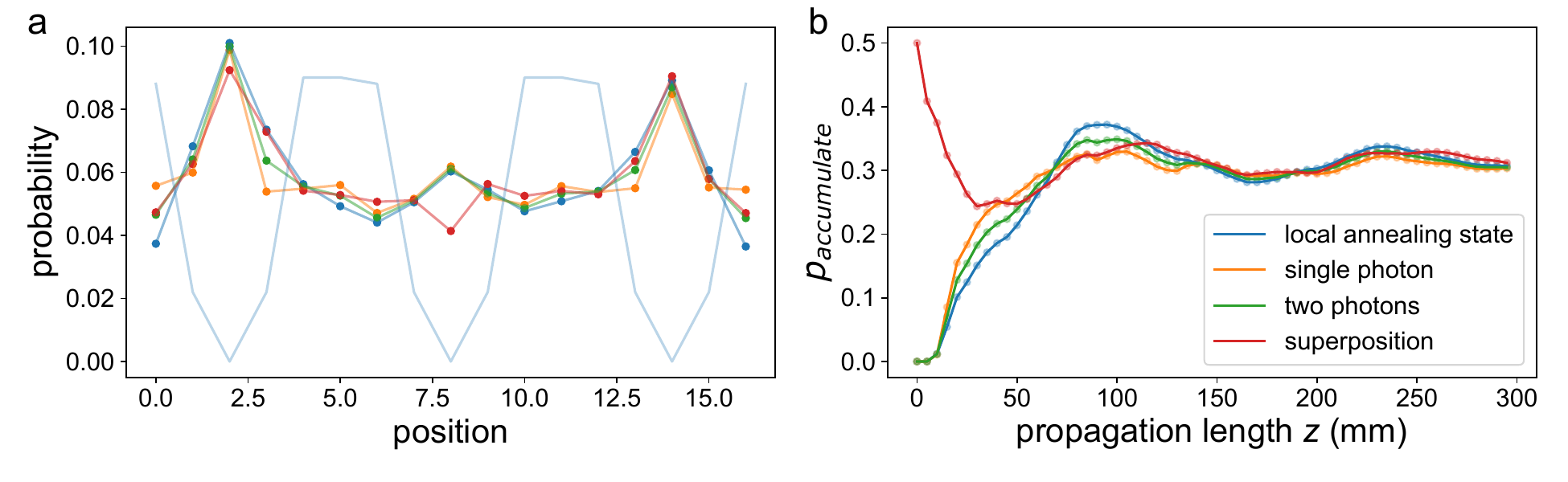}
    \caption{Search results of different injection states. \textbf{a} \red{For each input state, the position distribution is averaged over 60 propagation lengths from \SI{0}{mm} to \SI{295}{mm} in \SI{5}{mm} steps.} Annealing represents the injection state undergone a local annealing process; single injection represents single waveguide injection as the injection state without an annealing process; distinguishable photon is the combination injection of annealing and single injection while the two photon are nonidentical, and superposition represents the injestate that state is the superposition of local annealing state and single injection state. \textbf{b} \red{Running average of the occupation of the rightmost well (sites 12--16) up to each propagation length.}}
    \label{fig:A9}
\end{figure*}

\begin{figure*}
    \includegraphics[scale=0.53]{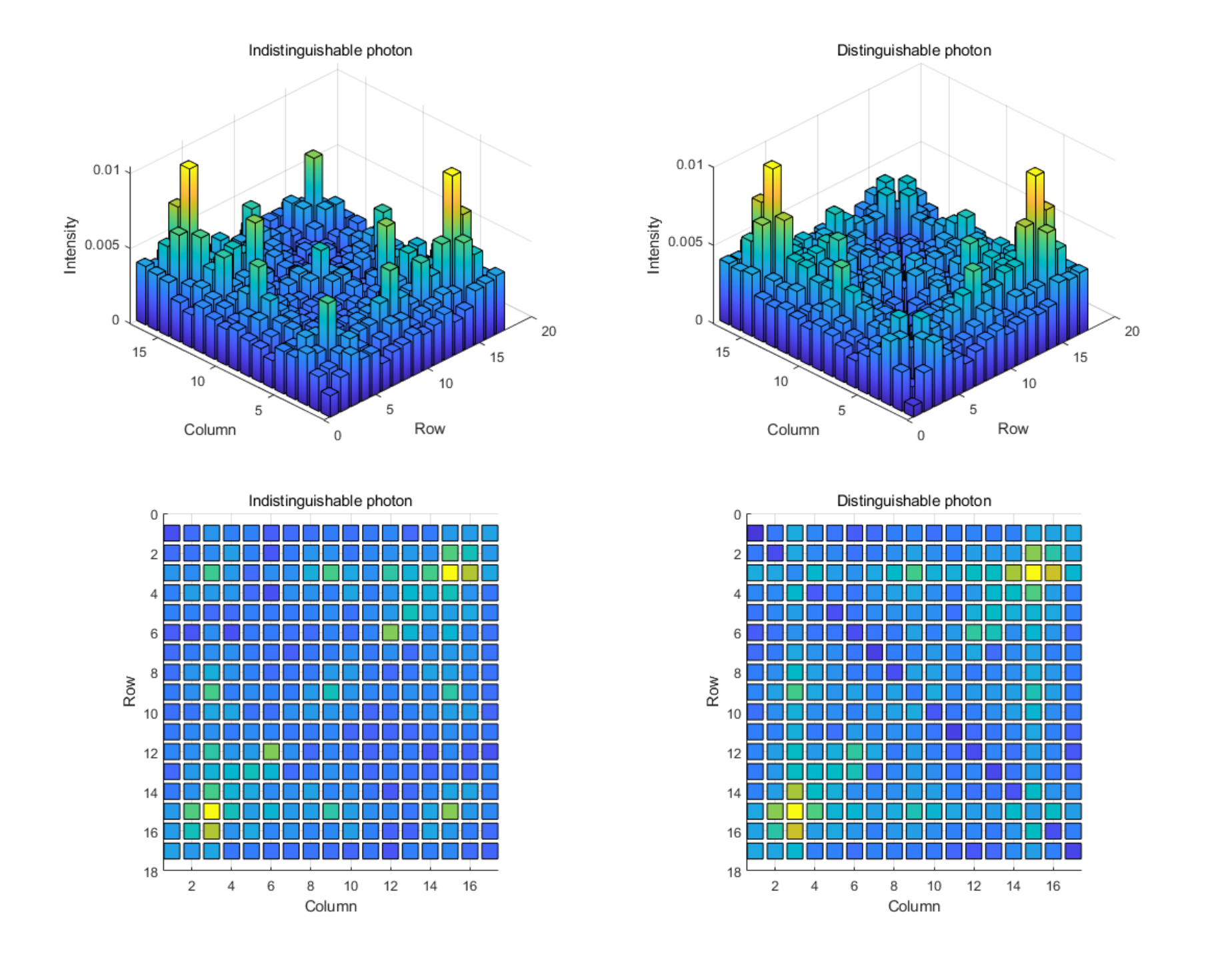}
    \caption{\red{Simulated two-photon coincidence distributions for indistinguishable and distinguishable inputs, averaged over propagation length in the three-well landscape of \fig{A4}.}}
    \label{fig:A10}
\end{figure*}

\begin{figure*}
    \centering
    \includegraphics[width=\linewidth]{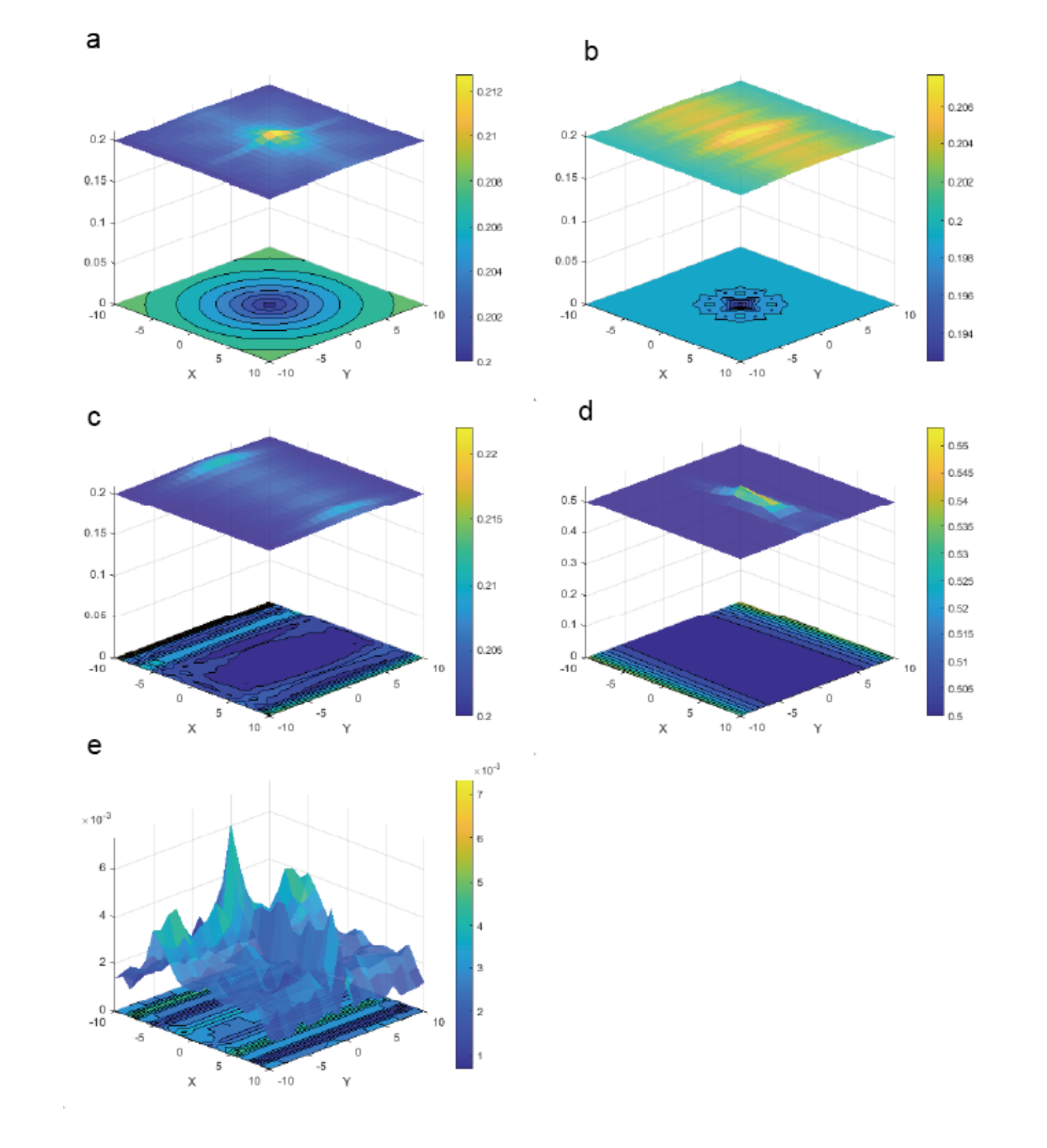}
    \caption{Quantum tunneling search in landscapes of exotic functions. From \textbf{a} to \textbf{e} are, in order, the Ackley function, the Drop-Wave function, the Levy function, the Dixon-Price function and the Michalewicz function.}
    \label{fig:A11}
\end{figure*}

\begin{figure*}
    \centering
    \includegraphics[width=\linewidth]{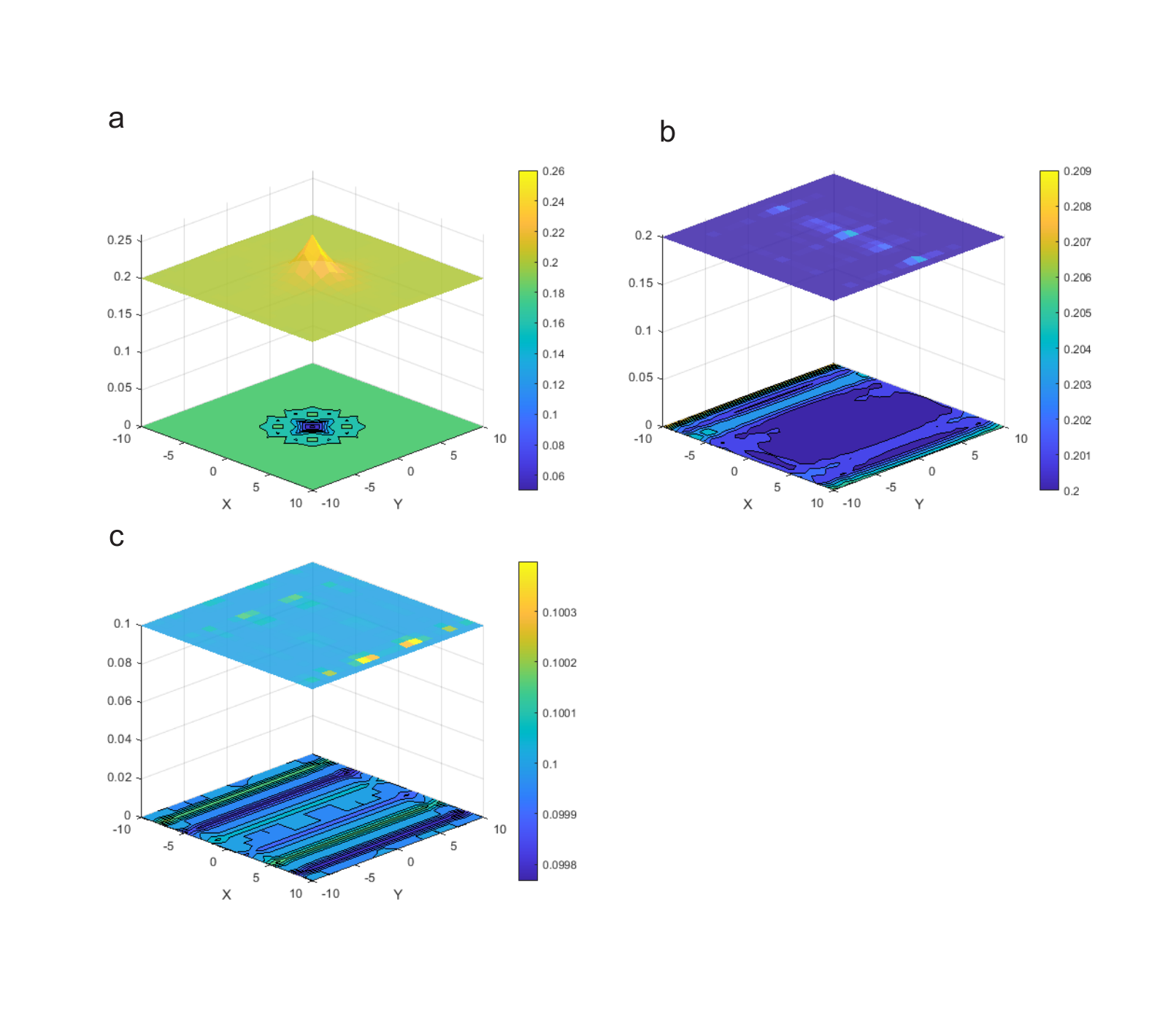}
    \caption{Quantum tunneling search in landscapes of exotic functions. From \textbf{a} to \textbf{c} are, in order, the Drop-Wave function, the Levy function, and the Michalewicz function.}
    \label{fig:A12}
\end{figure*}

\begin{figure*}
    \centering
    \includegraphics[scale=0.4]{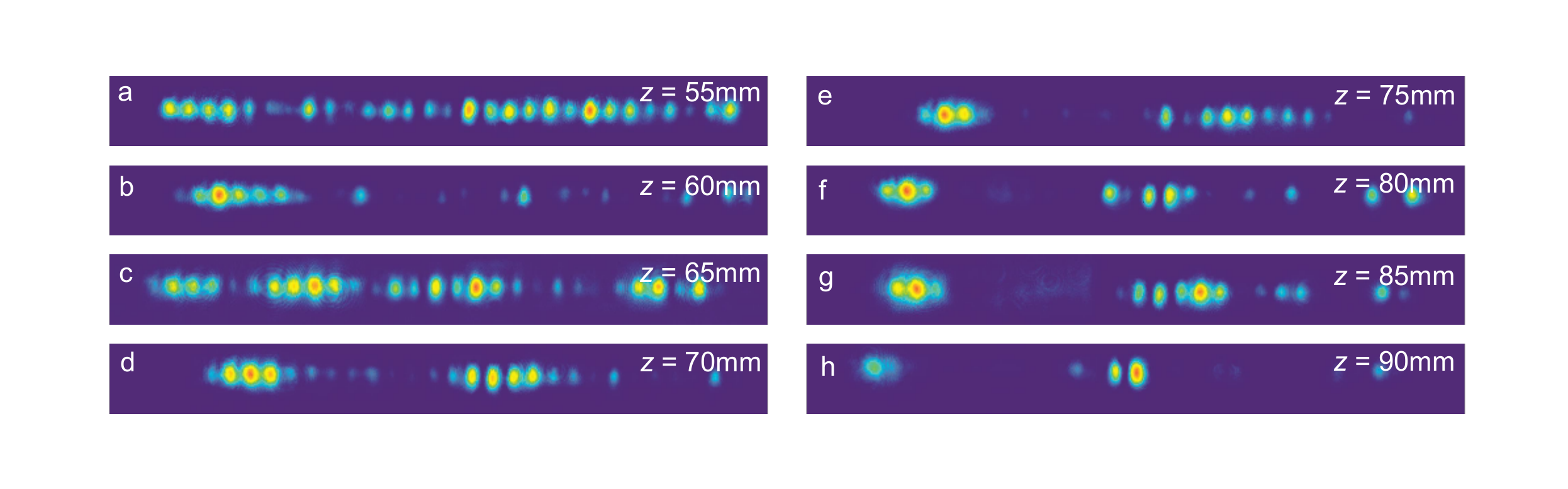}
    \caption{{Experimental global minimum quantum tunneling search with local annealing and varying Hamiltonian in a landscape with uneven local minima. Please refer to Figure 5 in the main text for the landscape.}}
\end{figure*}

\end{document}